\documentclass{aa}  

\usepackage{supertabular}  
\usepackage{graphicx}
\usepackage{txfonts}
\usepackage[dvipsnames]{xcolor}

\newcommand{\msun}{\ensuremath{\rm M_{\odot}}}
\newcommand{\rsun}{\ensuremath{\rm R_{\odot}}}

\begin{document}

   \title{AT\,2021hdr: A candidate tidal disruption of a gas cloud by a binary super massive black hole system}


\titlerunning{AT\,2021hdr}
\authorrunning{Hernández-García et al.}

   \author{L. Hernández-García
          \inst{1,2,3},
          A. M. Muñoz-Arancibia\inst{1,4},
          P. Lira\inst{5,2},
          G. Bruni\inst{6},
          J. Cuadra\inst{7,2},
          P. Arévalo\inst{3,2},
          P. Sánchez-Sáez\inst{8},
          S. Bernal\inst{3,2},
          F.E. Bauer\inst{9,10,1,11},
          M. Catelan\inst{9,10,1},         
          F. Panessa\inst{6},
          M. Pávez-Herrera\inst{9},
          C. Ricci\inst{12,13},
          I. Reyes-Jainaga\inst{14}, 
          B. Ailawadhi\inst{15,16},
          V. Chavushyan\inst{17,18},
          R. Dastidar\inst{1,19},
          A. Deconto-Machado\inst{20}, 
          F. Förster\inst{21,1,4,5},
          A. Gangopadhyay\inst{22},
          A. García-Pérez\inst{17,23},
          I. Márquez\inst{20},
          J. Masegosa\inst{20},
          K. Misra\inst{15},
          V.M Patiño-Alvarez\inst{17,24},
          M. Puig-Subirà\inst{20},
          J. Rodi\inst{6},
          \and
           M. Singh\inst{25}
          }

   \institute{Millennium Institute of Astrophysics (MAS), Nuncio Monseñor Sótero Sanz 100, Providencia, Santiago, Chile\\
              \email{lorena.hernandez@uv.cl}
        \and Millennium Nucleus on Transversal Research and Technology to Explore Supermassive Black Holes (TITANS) 
         \and
             Instituto de F\'isica y Astronom\'ia, Facultad de Ciencias,Universidad de Valpara\'iso, Gran Bretana 1111, Playa Ancha, Valpara\'iso, Chile
        \and Center for Mathematical Modeling, Universidad de Chile, Beauchef 851, Santiago 8370456, Chile 
        \and Departamento de Astronomía, Universidad de Chile, Casilla 36D, Santiago, Chile
        \and INAF – Istituto di Astrofisica e Planetologia Spaziali, Via del Fosso del Cavaliere 100, Roma, 00133, Italy
        \and Departamento de Ciencias, Facultad de Artes Liberales, Universidad Adolfo Ibáñez, Av.\ Padre Hurtado 750, Viña del Mar, Chile
        \and European Southern Observatory, Karl-Schwarzschild-Strasse 2, 85748 Garching bei München, Germany
        \and Instituto de Astrof{\'{\i}}sica, Facultad de F{\'{i}}sica, Pontificia Universidad Cat{\'{o}}lica de Chile, Campus San Joaquín, Av. Vicuña Mackenna 4860, Macul Santiago, Chile, 7820436
        \and Centro de Astroingenier{\'{\i}}a, Facultad de F{\'{i}}sica, Pontificia Universidad Cat{\'{o}}lica de Chile, Campus San Joaquín, Av. Vicuña Mackenna 4860, Macul Santiago, Chile, 7820436
        \and Space Science Institute, 4750 Walnut Street, Suite 205, Boulder, Colorado 80301
        \and Núcleo de Astronomía de la Facultad de Ingeniería, Universidad Diego Portales, Av. Ejército Libertador 441, Santiago, Chile
        \and Kavli Institute for Astronomy and Astrophysics, Peking University, Beijing 100871, People’s Republic of China
        \and Data Observatory Foundation, Santiago, Chile
        \and Aryabhatta Research Institute of Observational Sciences, Manora Peak, Nainital 263 001, India
        \and Department of Physics, Deen Dayal Upadhyaya Gorakhpur University, Gorakhpur-273009, India
        \and Instituto Nacional de Astrofísica, Óptica y Electrónica, Luis Enrique Erro 1, Tonantzintla, Puebla 72840, México
        \and Center for Astrophysics | Harvard \& Smithsonian, 60 Garden Street, Cambridge, MA 02138, USA
        \and Instituto de Astrofísica, Universidad Andres Bello, Fernandez Concha 700, Las Condes, Santiago RM, Chile
        \and Instituto de Astrofísica de Andalucía - CSIC, Glorieta de la Astronomía s/n, 18008 Granada, Spain
        \and Data and Artificial Intelligence Initiative (IDIA), Faculty of Physical and Mathematical Sciences, Universidad de Chile, Chile
        \and Department of Astronomy, The Oskar Klein Center, Stockholm University, AlbaNova 106 91, Stockholm, Sweden
        \and Dipartimento di Fisica, Università degli Studi di Torino, via Pietro Giuria 1, I-10125 Torino, Italy
        \and Max-Planck-Institut f\"ur Radioastronomie, Auf dem Hügel 69, D-53121 Bonn, Germany
        \and Indian Institute of Astrophysics, Koramangala 2nd Block, Bangalore 560034, India
             }

   \date{Received June 28, 2024; accepted October 11, 2024}

 
  \abstract{
   With a growing number of facilities able to monitor the entire sky and produce light curves with a cadence of days, in recent years there has been an increased rate of detection of sources whose variability deviates from standard behavior, revealing a variety of exotic nuclear transients. 
   The aim of the present study is to disentangle the nature of the transient AT\,2021hdr, whose optical light curve used to be consistent with a classic Seyfert 1 nucleus, which was also confirmed by its optical spectrum and high-energy properties.
   From late 2021, AT\,2021hdr started to present sudden brightening episodes in the form of oscillating peaks in the Zwicky Transient Facility (ZTF) alert stream, and the same shape is observed in X-rays and UV from \textit{Swift} data. The oscillations occur every $\sim$60-90 days with amplitudes of $\sim$0.2 mag in the g and r bands.
   Very Long Baseline Array (VLBA) observations show no radio emission at milliarcseconds scale.
   It is argued that these findings are inconsistent with a standard tidal disruption event (TDE), a binary supermassive black hole (BSMBH), or a changing-look active galactic nucleus (AGN); neither does this object resemble previous observed AGN flares, and disk or jet instabilities are an unlikely scenario. Here, we propose that the behavior of  AT\,2021hdr might be due to the tidal disruption of a gas cloud by a BSMBH.
   In this scenario, we estimate that the putative binary has a separation of $\sim$0.83 mpc and would merge in  $\sim$7$\times$10$^4$ years.
   This galaxy is located at 9 kpc from
a companion galaxy, and in this work we report this merger for the first time. The oscillations are not related to the companion galaxy.
   }


   \maketitle
%

\section{Introduction}

The centers of most galaxies are thought to host a supermassive black hole (SMBH).
If fed by a surrounding accretion disk, the system shines and becomes an active galactic nucleus
\citep[AGN;][]{rees1984}.
A common property of AGN is their variability across the entire electromagnetic spectrum \citep{netzer2013}, which occurs stochastically and is thought to be related mostly to flux variations in the accretion disc, with typical amplitudes of   $<$ 0.5 mag in the optical continuum on timescales of between months and years \citep[e.g.,][]{vanden2004, macleod2016}.

In recent decades, objects whose behavior deviates from typical AGN-like variations have been discovered. These include tidal disruption events (TDEs, \citealt{gezari2021, zabludoff2021}), binary supermassive black hole (BSMBH; \citealt{charisi2016, derosa2019})  candidates, changing-look (CL; \citealt{lamassa2015, ricci2023, zeltyn2024})  events, the recently discovered ambiguous nuclear transients \citep[ANTs; e.g.,][]{hinkle2022}, and anomalous-variability flares, which are due to disk instabilities or similar perturbations to the accretion process \citep{graham2017, Trakhtenbrot2019NA, frederick2021, dotti2023}.

 Many of the previous novel SMBH-related events were found thanks to the ever-increasing expansion of optical time domain facilities, which monitor the entire observable sky in search of variable objects. Among them, the \textit{Zwicky} Transient Facility \citep[ZTF,][]{graham2019, bellm2019, masci2019} observes the same region of the sky approximately every 3 days, allowing the construction of well-sampled light curves, and alerting the scientific community to significant changes in the sky.
A multiwavelength perspective is often crucial to properly assess the evolving emission mechanisms and energetics of these variable sources \citep[e.g.,][]{lore2023}.
 For this, the Neil Gehrels \textit{Swift} Observatory \citep{2005SSRv..120...95R, 2005SSRv..120..165B} is an ideal facility, as it allows a rapid response to proposals that request follow-up observations, obtaining data in X-rays, UV, and optical.

In this paper, we present an intriguing source that was discovered in 2021 thanks to the Automatic Learning for the Rapid Classification of Events broker \citep[ALeRCE;][]{forster2021} using ZTF data, and that we began monitoring with \textit{Swift} in 2022. We complement these data with observations in radio and optical spectra.
The host of this new transient or variable event, whose light curves show relatively well-defined oscillations on timescales of months, is an early stage merger composed of an AGN and a low-ionization nuclear emission line region (LINER); the combination of these two rare phenomena is reported for the first time \citep{derosa2019}.

The paper is organized as follows: in Sect. \ref{selection} we present the selection of the target, in Sect. \ref{host} we include information about the nuclear region of the host galaxy and its environment, in Sect. \ref{datamain} we present the data used for our analysis,
in Sect. \ref{moonnitoring} we describe the multiwavelength light curves obtained during the transient phase, and in Sect. \ref{disc} we discuss the main results obtained in this work. Finally, we present our conclusions in Sect. \ref{summary}. The appendix provides further details and analysis of the data as well as the full light curves.

\section{Target selection \label{selection}}

    \begin{figure*}
\sidecaption
   \includegraphics[width=12cm]{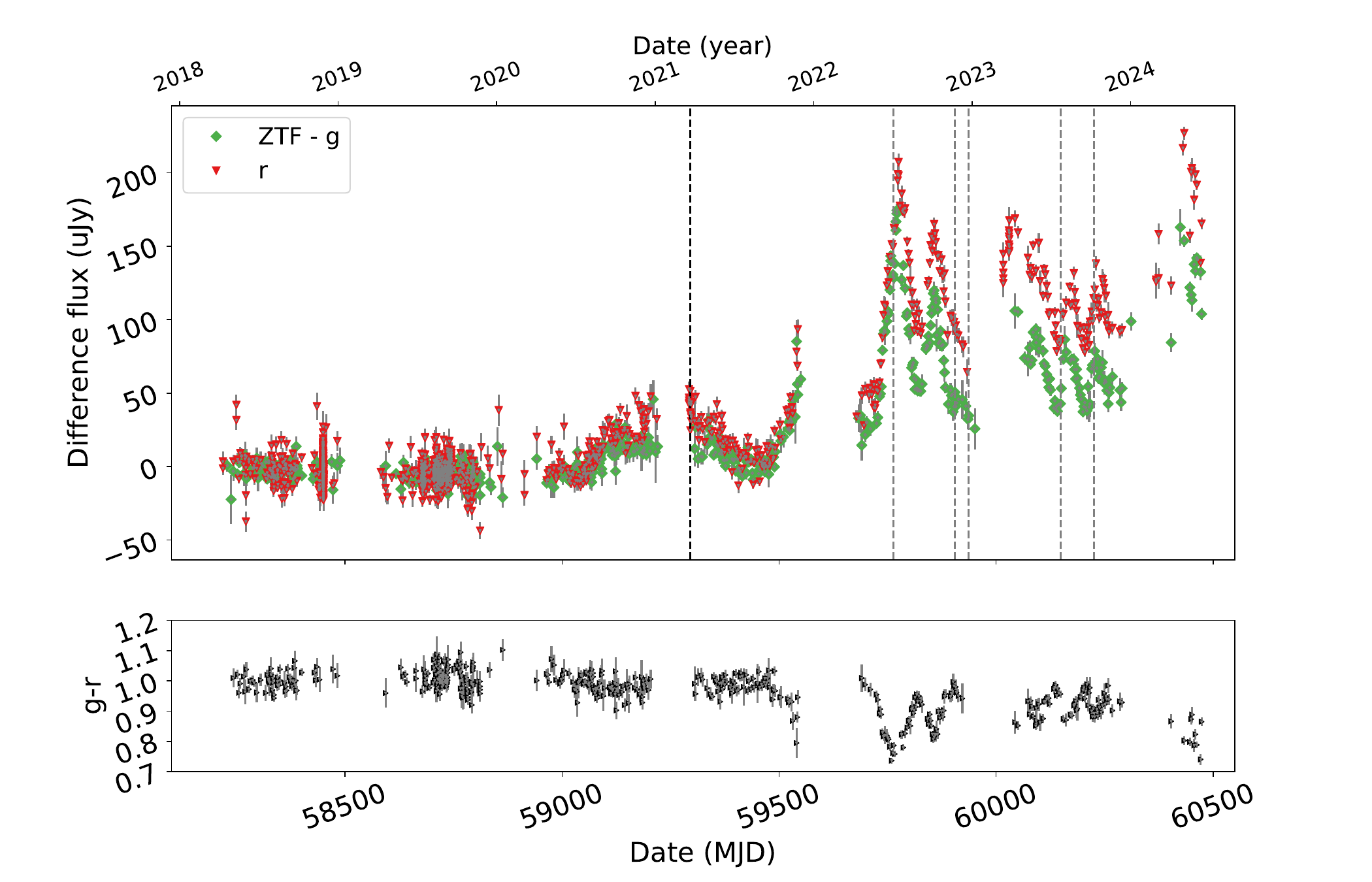}
      \caption{ZTF light curve of AT\,2021hdr between 2018-2024. (Top panel): ZTF light curve (in difference flux from PSF forced photometry) of AT\,2021hdr in the g (green squares) and r (red triangles).  The black dashed line represents the date of the first ZTF alert, and the gray dashed lines correspond to dates when optical spectra were obtained. (Bottom panel): ZTF g-r color evolution in total magnitude (see text for details).
              }
         \label{ZTF_LC}
   \end{figure*}


%
\begin{table}
\caption{Re-brightenings observed in the ZTF light curve.}             
\label{table:rebright}      
\centering                          
\begin{tabular}{l c c}        
\hline\hline                 

Starting  &   Peak apparent        &       Amplitude \\
date & magnitude & $\Delta m_{\rm g, AB, peak}$ \\
(dd-mm-yy) &(mag)& (mag)\\
\hline
02-11-21 & 17.57/- & 0.26/- \\
28-05-22 & 17.36/16.64 & 0.43/0.16 \\
31-08-22 & 17.73/16.69 & 0.27/0.15 \\
24-12-22$^*$ & 17.53/16.66 & 0.27/0.15 \\
14-07-23 & 17.78/16.74 & 0.18/0.07 \\
13-09-23 & 17.79/16.72 & 0.15/0.07 \\
\hline  
\end{tabular}
\tablefoot{The columns include the starting date of the re-brightening, its peak apparent magnitude, and amplitude in the g/r bands. $^*$It is uncertain whether there are one or two oscillations after this date.}
\end{table}

\subsection{AT\,2021hdr}

AT\,2021hdr / ZTF21aaqqwsa was discovered on March 22, 2021, in the ZTF public alert stream at Right Ascension (RA) = 321.00142\textdegree, Declination (Dec.) = 34.15319\textdegree (J2000), and was reported to the Transient Name Server (TNS) by the ALeRCE broker \citep{forster2021} as a supernova candidate on March 26, 2021 (TNS \#104012, \citealt{discovreport}). AT 2021hdr / ZTF21aaqqwsa has shown a remarkable variability pattern with multiple oscillating peaks \citep{MunozArancibia2024}. 
Figure \ref{ZTF_LC} shows the ZTF light curve obtained from the Forced Photometry Service in difference flux \citep[][see Sect. \ref{dataztf} for details]{masci2023}.
After the first alert in the ZTF alert stream, corresponding to the vertical dashed line in Fig. \ref{ZTF_LC}, the source has experienced five re-brightenings in the ZTF data. These can be found in Table \ref{table:rebright}.

The bottom panel of the figure shows the ZTF g-r color evolution with time. We estimated the g-r color per g-band epoch by subtracting the closest r-band apparent magnitude (within the same night) from each g-band apparent magnitude. This method is limited to nights where AT\,2021hdr photometry was measured in both filters, thus avoiding color estimates for nights that do not meet this criterion. We note that this color estimate includes the host-galaxy contribution to the brightness, as well as dust attenuation from both host-galaxy and Milky Way dust.

\subsection{Host and environment \label{host}}

The host galaxy of AT\,2021hdr, 2MASX J21240027+3409114 \citep{2masx2006}, was classified as a Seyfert nucleus by \cite{parisi2014} based on a spectrum obtained using the San Pedro Martir (SPM) Telescope ($z{=}0.083$). Also known as PBC\,J2123.9+3407, it was detected for the first time at X-rays in the Palermo \textit{Swift}-BAT hard X-ray catalog \citep{cusumano2010}. 
Its hard X-ray properties are described in the \textit{Swift} BAT 157-Month Hard X-ray Survey; its spectrum has a spectral index of 2.41[1.69-3.51], resulting in a luminosity of logL = 44.0 erg s$^{-1}$ in the 14--195 keV energy band \citep[][see Appendix \ref{swiftbat}]{lien2023}.

\begin{figure*}
   \centering
   \includegraphics[width=16cm]{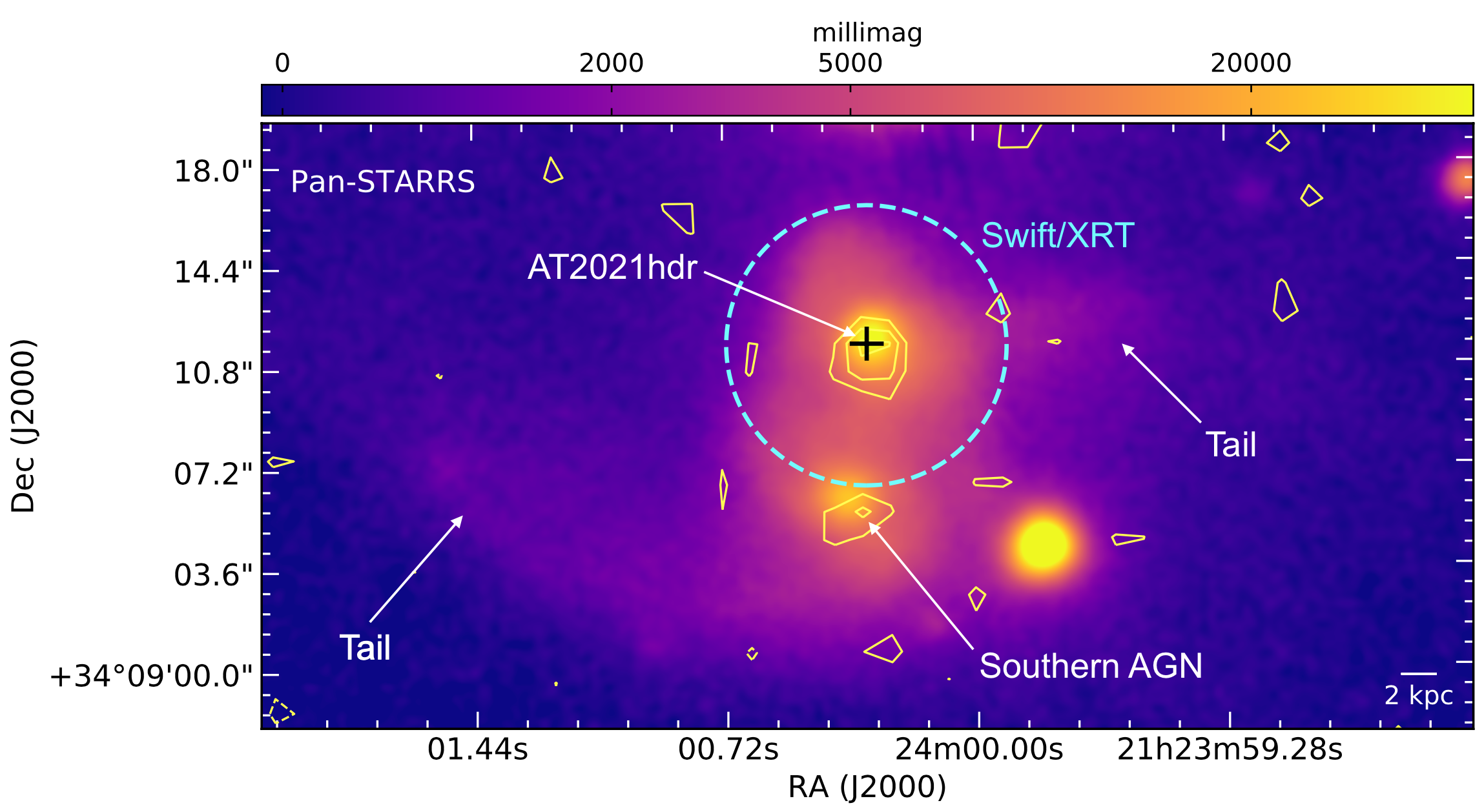}
      \caption{Pan-STARRS stacked i-band image of the AT\,2021hdr (black cross) host and environment. The cyan circle indicates the {\textit{Swift}}/XRT source position error (90\%). VLASS contours at 2$\sigma$, 3$\sigma$, and 4$\sigma$ are overlaid in yellow. The locations of the secondary AGN and tidal tails are indicated.
              }
         \label{panstarrs}
   \end{figure*}

The host of AT\,2021hdr is clearly detected in Pan-STARRS (see Fig. \ref{panstarrs}), and appears to be part of an ongoing early-stage merger with strong tidal tails. The second object, 2MASS J21240037+3409058, at 6" or 9 kpc to the south, did not have a redshift estimation. We obtained a spectrum with SPM on November 23, 2022, that allowed us to measure a redshift of $z{=}0.081$ and classify it as a LINER nucleus (see Appendix \ref{analysisoptical}). Thus, the system is a newly confirmed galaxy merger reported here for the first time.

In the radio band, the host galaxy is detected at 4$\sigma$ significance at 3 GHz at kiloparsec (kpc) scale (see also following section).
When observed at parsec (pc) scale, no detection is found at a $\sim70\, \mu\rm{Jy/beam}$ flux density level, implying that there is no strong jet or core emission ($L<10^{37.5}$ erg/sec at 5 GHz). Thus, the radio emission observed at kiloparsec scales does not appear to be related to the transient in AT\,2021hdr, but rather to extended emission such as star-forming or wind activity (see Appendix \ref{radioemission}).

\section{\label{datamain}Data}

\subsection{Radio}

\subsubsection{Very Large Array Sky Survey}

The galaxy hosting AT\,2021hdr is in the footprint of the Very Large Array Sky Survey \citep[VLASS, 2-4 GHz, $\sim$3" FWHM,][]{lacy2020}. 
It was observed twice: on June 5, 2019, and October 13, 2021 (a 2.3 yr span).
The images show a detection in both epochs at a significance level of 4$\sigma$, with a flux density of $\sim$550 $\mu$Jy (see Table \ref{tab:VLASS}). Flux density values are consistent within errors. 
However, the highest variations in the optical band are after 2022, and so we cannot exclude a radio flux increase during the past year. 

The southern galaxy is also detected in both epochs, with a flux density of $\sim$400 $\mu$Jy (402$\pm$144 $\mu$Jy in the first epoch, and 400$\pm$122 $\mu$Jy in the second one). The radio 3$\sigma$ and 4$\sigma$ contours encompass the bulges of the corresponding host galaxies, as seen in the optical PanSTARRS images (see Fig. \ref{panstarrs}). A tentative Gaussian fit for the radio emission co-located with AT\,2021hdr results in a deconvolved size of smaller than the beam, suggesting the presence of a compact component on the kpc scale (one beam of 3" corresponds to $\sim$5 kpc at a redshift of 0.083). However, given the low flux density and statistical significance, this should be considered only a rough estimate. Indeed, extended emission with a lower surface brightness could still be present below the noise level. 


\begin{table}[]
    \centering
    \caption{Values extracted from the two VLASS survey epochs.}
    \begin{tabular}{cccccc}
    \hline
        Survey      &  Date         &   RMS &  FWHM &   Flux density   \\
          VLASS      &  ddmmyyyy    &   ($\mu$Jy/beam) & ("$\times$") & $\mu$Jy \\
    \hline
       Ep.1  &  05-06-2019   & 138   & 2.5$\times$2.2  & 569$\pm$149     \\
       Ep.2  &  13-10-2021   & 115   & 2.6$\times$2.2  & 543$\pm$127     \\
    \hline
    \end{tabular}
    \label{tab:VLASS}
\end{table}


%

%

\subsubsection{Very Long Baseline Array }

Observations with the Very Long Baseline Array (VLBA) were performed on June 29, 2023, through an approved DDT proposal (Project ID: BH241) to test the presence of a compact radio core and a possible jet in both nuclei. Observations were carried out at 5 GHz and 8 GHz (C- and X-band, respectively), and data correlated when applying two different phase centers to include the location of AT\,2021hdr and the southern AGN as well. Data were reduced with the Astronomical  Image  Processing  System  ({\tt{AIPS}}\footnote{\url{http://www.aips.nrao.edu}}) following standard procedures. Imaging was performed in {\tt{CASA}}\footnote{\url{https://casa.nrao.edu/index.shtml}}.
We were able to reach an RMS of 23 $\mu$Jy/beam for both nuclei at 5 GHz, and a slightly higher RMS of 30 $\mu$Jy/beam at 8 GHz. The angular resolution was 5$\times$4 milliarcsec at 5 GHz and 2$\times$1 milliarcsec at 8 GHz. No emission was detected at the location of AT\,2021hdr or its companion galaxy, corresponding to an upper limit of $10^{37.5}$ erg/sec at 5 GHz for AT\,2021hdr.

\subsection{Optical}

\subsubsection{\label{dataztf}Zwicky Transient Facility }

The ZTF \citep{graham2019, bellm2019, masci2019, masci2023} has been surveying the northern sky every three days in the g, r, and i optical filters since 2018 
and offers different services to the facilities described above, including (1) a public alert system, for real-time time-domain science,
(2) data releases (DRs) every two months, including photometry measurements on the science images, 
and 
(3) a Forced Photometry Service on demand and per source, including photometry measurements on the reference-subtracted science images.                                                                        

For an alert to be generated, a source has to show a variation above a 5$\sigma$  confidence level with respect to a reference image\footnote{A reference image is generated by ZTF from the stacking of at least 10 images of each field.} .
AT\,2021hdr triggered its first alert on March 22, 2021, and was reported by the ALeRCE broker to the TNS \citep{discovreport}. The alert light curve can be seen in the ALeRCE Web Interface\footnote{\url{https://alerce.online/object/ZTF21aaqqwsa}} provided by the ALeRCE broker \citep[ALeRCE,][]{forster2021}, with ZTF ID ZTF21aaqqwsa.

For this work, we retrieved data from the ZTF Forced Photometry Service. The measurements obtained from this service are less affected by extranuclear emission than in DRs because they are obtained from the reference-subtracted images, therefore isolating the variable nuclear component.
The light curves were constructed with the criteria explained in \cite{masci2023} and in \cite{lore2023}, including quality filtering and rejecting bad-data quality flags. We did not apply color correction to the data because AT\,2120hdr shows color variations (see Fig. \ref{ZTF_LC}).

The difference PSF magnitudes were converted into apparent magnitude following \cite{forster2021}. 
In this manuscript, we present the ZTF light curve of AT\,2021hdr in difference flux (i.e., the variable flux) in Figs. \ref{ZTF_LC} and \ref{fig_tdefittingtotal}, and in apparent magnitude (i.e., the magnitude corrected for the contribution of the host galaxy as measured in the reference image) in Fig. \ref{fig_LCs}.

\subsubsection{\label{optspectra}Spectra}

\cite{parisi2014} presented the first spectrum of this nucleus obtained using the 2.1 m telescope of the OAN in San Pedro M\'artir (SPM), M\'exico, in 2010.
AT\,2021hdr was later observed with the Liverpool Telescope (LT) on July 3, 2022, and was classified as an AGN by \cite{classificationreport2022}. We obtained a copy of the reduced 1D spectrum from TNS. 

In the present work, we analyze new spectra obtained from different observatories, as follows.
The source was observed with the SPM on November 22, 2022, and October 9, 2023, using a Boller and Chivens spectrograph. A spectrum of the nucleus of the southern galaxy located at 6 arcsec from AT\,2021hdr was obtained on November 23, 2022. 
The instrument was equipped with a 2K$\times$2K pixels E2V 4240 CCD. For flux calibration, a standard star was observed every night.
Wavelength calibration was done using CuHeNeAr comparison lamps. A slit width of 2.5 arcsec was used, resulting in a spectral resolution of 10 \AA. 
 

An optical spectrum was acquired with the Hanle Faint Object Spectrograph and Camera \citep[HFOSC][]{Prabhu2010} mounted on the 2.01 m Himalayan Chandra Telescope (HCT) at IAO, Hanle, on December 23, 2022. For the observations, a grism Gr7 (3800–7800 $\AA$) was used with a resolution of 1330 and dispersion of 1.45 $\AA$ pixel$^{-1}$.
The slit width was 0.77 arcsec and the exposure time was 3600 s.

Spectral data were taken with the Andalucian Faint Object Spectrograph and Camera (ALFOSC) attached to the 2.5 m Nordic Optical Telescope (NOT\footnote{Roque de los Muchachos Observatory, La Palma, Canary Islands, Spain}) on July 23, 2023. Grating \#4 was used, which covers the spectral range 3200A-9600\AA, resulting in a spectral sampling of 3.3 $\AA$ pixel$^{-1}$. A slit of 1 arcsec was used. A total on-target exposure time of 3600 seconds was gathered in three exposures taken for cosmic rays and bad pixel removal. 
Arc lamp exposures were obtained before and after each target observation. A standard star was observed for flux calibration with a 10 arcsec width slit.

Spectroscopic data reduction was carried out using IRAF, following the standard steps of bias subtraction, flat-field correction, wavelength calibration, atmospheric extinction correction, and flux calibration. The sky background level was determined by taking median averages over two strips on both sides of the galaxy signal, and subtracting it from the final combined galaxy spectrum. 

\subsection{UV and X-rays}


The Ultraviolet and Optical Telescope \citep[UVOT,][]{2005SSRv..120...95R} on board the Neil Gehrels \textit{Swift} Observatory has six primary photometric filters: V (centered at 5468 \AA), B (at 4392 \AA), U (at 3465 \AA), UVW1 (at 2600 \AA), UVM2 (at 2246 \AA) and UVW2 (at 1928 \AA). Archival data from 2020 included only UVW2 data, whereas we obtained observations through target of opportunity (ToO) proposals in November 2022 in all six filters, between July and December 2023 in the UVW1, UVW2, and UVM2 filters, and from December 2023 until April 2024 in the UVM2 filter only. Data have been obtained in the UVM2 filter since April 2024 through a regular GI proposal \#2023221.
 
The {\sc uvotsource} task within software HEASoft version 6.30 was used to perform aperture photometry using a circular aperture of radius 5 arcsec centered on the coordinates of AT\,2021hdr. A background region free of sources  was selected by adopting a circular region of 20 arcsec close to the nucleus. 

An issue with UVOT arose whereby some observations were affected by spacecraft jitter, which causes the sources in the UVOT images to appear elongated, and not point-like. When an observation suffers from this jitter, errors are induced in the standard photometry methods, underestimating magnitudes by 0.1-0.3 mag. The UVOT team advised that the data can be used but with caution (private communication). This effect was observed between August 2023 and the start of April 2024 \citep{2023GCNswift, 2024GCNswift}.

Simultaneous to the UVOT data, observations with the \textit{Swift} X-ray Telescope \citep[XRT,][]{2005SSRv..120..165B}, also on board the Neil Gehrels \textit{Swift} Observatory, were taken in the Photon Counting mode through the ToO proposals and the regular GI proposal \#2023221.
The data reduction  was performed following standard routines described by the UK Swift Science Data Centre (UKSSDC) using the software in HEASoft version 6.30.1. Calibrated event files were produced using the routine {\sc xrtpipeline}, accounting for bad pixels and effects of vignetting, and exposure maps were also created. Source and background spectra were extracted from circular regions with 20 arcsec and 50 arcsec radius, respectively. The {\sc xrtmkarftask} was used to create the corresponding ancillary response files. The response matrix files were obtained from the HEASARC CALibration DataBase. The spectra were grouped to have a minimum of 20 counts per bin using the {\sc grppha} task. The light curve in Fig. \ref{fig_LCs} is presented in counts per second to have model-independent measurements.

To carry out a spectral analysis of the X-ray data, we used XSPEC v.12.10.1. We assumed a Galactic absorption of $N_{\rm Gal} = 1.19\times 10^{21}$ cm$^{-2}$\citep{1990ARAA..28..215D}. Errors correspond to 90\% confidence limits.

All individual spectra were fitted simultaneously with a power law including Galactic absorption. First, we fitted the spectra with frozen parameters assuming standard values, but this resulted in a $\chi ^2$ of 1091.3 for 512 degrees of freedom (dof). We then left the spectral index and normalization parameters free, both independently and together. performing spectral fittings each time to determine the best fit.
The model forcing the same spectral index for all data sets and a varying normalization resulted in a best-fit value of  $\Gamma$ = 1.4$^{+0.6}_{-0.5}$ for $\chi ^2$/dof = 389.58/468 = 0.83. All other models resulted in $\chi ^2$/dof $>$ 1.6. Adopting an intrinsic absorption did not improve any of the fitting results.
From this model, we obtained fluxes in the 0.5-10 keV energy band in the range $7.6\times 10^{-13} - 6.7\times10^{-12}$ erg cm$^{-2}$  s$^{-1}$, with a mean value of $2.7\times10^{-12}$ erg cm$^{-2}$  s$^{-1}$. At the redshift of the source, this corresponds to a mean luminosity of log\,L$_{0.5-10 {\rm keV}} = 43.6^{+0.4}_{-0.5}$ erg s$^{-1}$.

We also estimated the hardness ratio (HR) as (H - S)/(H + S) for each data set, with S being the count rate in the $0.5-1.0$ keV energy band, and H being the count rate in the $1.0-10$ keV energy band. We obtained values of between 0.3 and 1.0, with a mean value of HR $= 0.7$. All measurements quoted above are in agreement with AGN standard values \citep[e.g.,][]{panessa2006, brightman2011}.

\section{Results of the multiwavelength monitoring \label{moonnitoring}}

\begin{figure*}
\sidecaption
  \includegraphics[width=12cm]{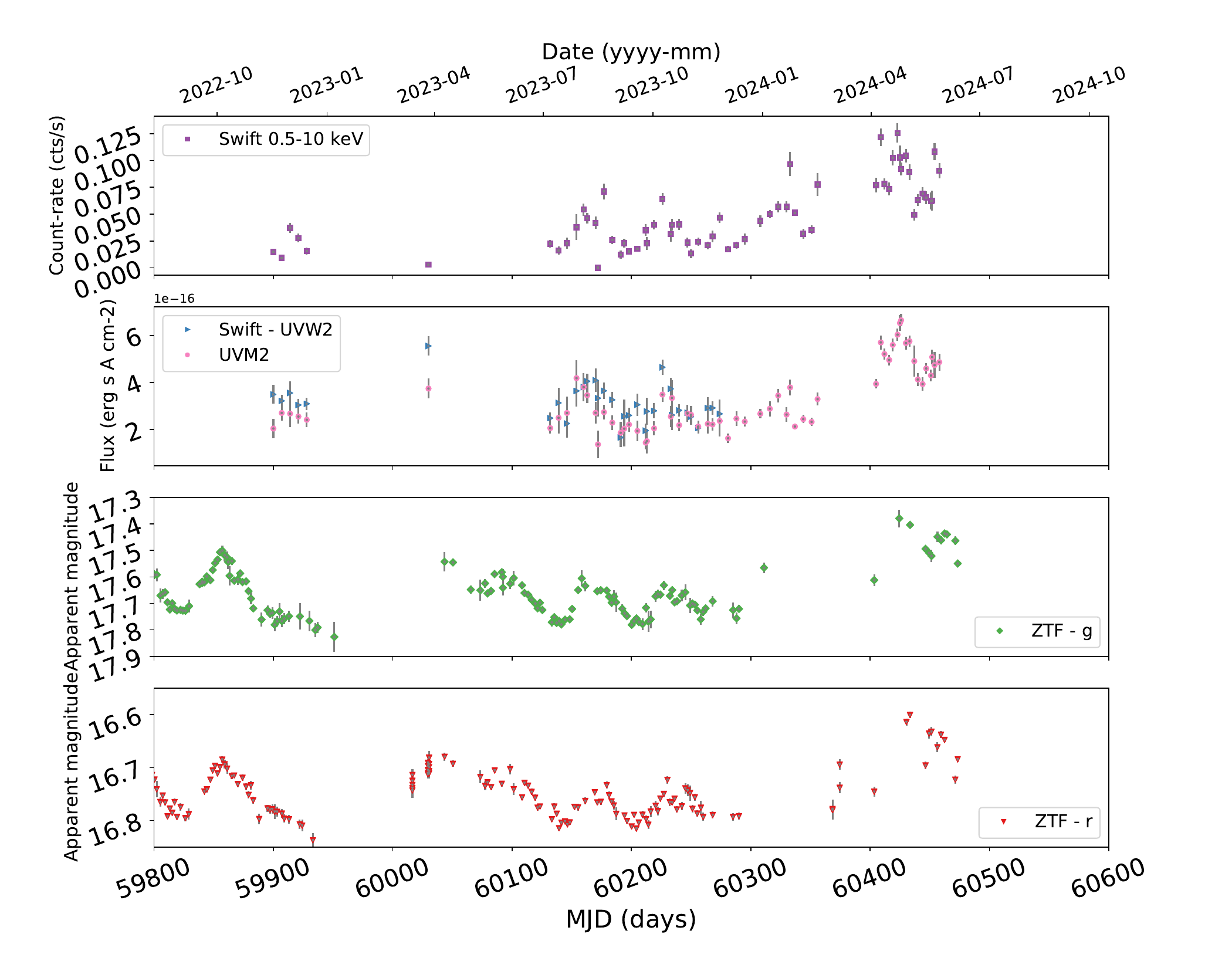}
     \caption{Light curves of AT\,2021hdr. From top to bottom: \textit{Swift}/XRT in the 0.5-10 keV energy band, \textit{Swift}/UVOT in the UVW2 (blue triangles) and UVM2 (pink circles), and ZTF in the g (green crosses) and r (red triangles) bands. Dates are between August 8, 2022, and March 31, 2024. }
     \label{fig_LCs}
\end{figure*}

The remarkable variability pattern of AT\,2021hdr is presented in Fig. \ref{fig_LCs}. The top panel shows \textit{Swift}/XRT data in the 0.5-10 keV energy band, taken between November 17, 2022, and June 14, 2024. An approximately weekly monitoring was started on July 7, 2023, with a gap in early 2024 when the source was behind the Sun.
The middle panel shows \textit{Swift}/UVOT data
taken simultaneously to the XRT data. After July 7, 2023, priority was given to observe with the UV filters, and therefore we present UVW2 and UVM2 measurements as optical observations were scarce. After December 3, 2023, data with the UVM2 filter only were taken. The bottom panel shows a zoom onto the ZTF Forced Photometry presented in Fig. \ref{ZTF_LC}. The observations have a cadence of $\sim 3$ days. Measurements correspond to optical apparent magnitude.

There are four oscillation episodes (numbered 1, 2, 5, and 6 in Fig. \ref{fig_tdefittingtotal}) for which we can roughly estimate both the rise and fall times in the g-band: 44 and 51 days (95 days in total and an amplitude of $\sim 0.4$ mag) between MJD 59727 and 59822; 35 and 52 days (87 days and amplitude $\sim 0.3$ mag) between MJD 59822 and 59909; 26 and 35 days (61 days and amplitude $\sim 0.2$ mag) between MJD 60139 and 60200; and 27 and 31 days (58 days and amplitude $\sim 0.15$ mag) between MJD 60200 and 60258. Thus, the rising times are shorter than the falling ones, but only for a few days, while the oscillation episode time span as well as their amplitude might be decreasing with time, although the behavior observed in 2024 appears rather different.

A similar behavior is observed in the X-rays and UV from \textit{Swift} data, although the cadence of the observations prevents us from making the same estimations. Time lags between the different bands cannot be estimated with the current available data.

   \begin{figure}
   \centering
   \includegraphics[width=9cm]{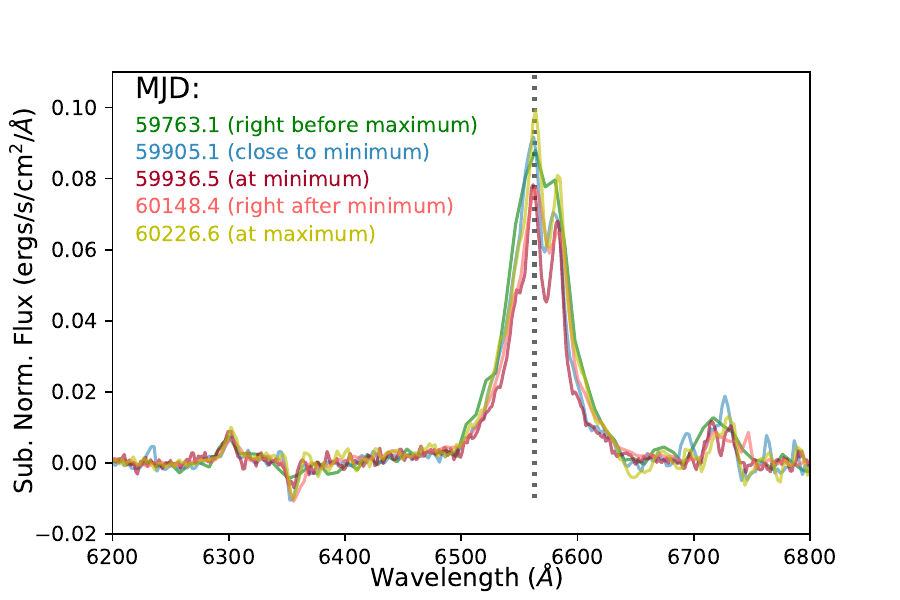}
      \caption{H$\alpha$ region of AT\,2021hdr. From top to bottom, spectra from the LT, SPM, HCT, NOT, and SPM are presented in chronological order. The spectra are normalized for visualization purposes (see text).
              }
         \label{fig_halpha}
   \end{figure}

As described in Sect. \ref{datamain}, archival optical spectra and new spectra were obtained by our group.
As the spectra were taken using different instruments, different slit settings, and different seeing conditions, allowing for different contributions of the significant host stellar light into the spectral aperture, a direct comparison between epochs is not possible, except for the SPM data obtained in 2022 and 2023 (see Appendix \ref{analysisoptical} for details). Nevertheless, Fig. \ref{fig_halpha} presents all data available to us around the H$\alpha$ spectral region. For this comparison, we normalized each spectrum so that the [O III]\,$\lambda5007$ line flux was equal, and subtracted a local continuum obtained by fitting a straight line to the flux measured from two small windows centered at 6225 and 6775 \AA. Besides the redshift correction, small shifts were also introduced to align the peak of the H$\alpha$, as this is dominated by the narrow component (see Appendix \ref{analysisoptical}). The result suggests small changes in the total flux of the H$\alpha$ emission line and very few changes in the profile of its broad component. In particular, the integrated fluxes between 6225 and 6775 \AA\ change by less than a factor 1.4.

From the SPM spectra, we estimated a black hole mass of M$_{BH} \sim$ 4$\times$10$^7$ M$_{\odot}$ (see Appendix \ref{analysisoptical}).
Using this mass, we can calculate the Eddington ratio, R$_{Edd}$. We estimated the X-ray luminosity in the 2-10 keV energy band for the spectra obtained in 2010 (pre-oscillations) and during the oscillations. From these luminosities, we calculated the bolometric luminosity of the source using the bolometric correction given by \cite{marconi2004}. We obtained luminosities of L$_{bol}$ = 1.71$\times$10$^{44}$ erg s$^{-1}$ and L$_{bol}$ = 3.38$\times$10$^{44}$ erg s$^{-1}$ for pre- and post-oscillations, respectively. From these measurements, we obtain R$_{Edd}$ $\sim$0.04 and $\sim$0.07 by dividing by the Eddington luminosity.
For comparison, we estimate a luminosity of L$_{bol}$ = 1.53$\times$10$^{44}$ erg s$^{-1}$ using the broad H${\alpha}$ emission line.

{\section{Source nature \label{disc}}}

The oscillations are observed in the optical, UV, and X-rays every approximately 60-90 days with amplitudes of $\sim$ 0.2 mag in the optical. The source oscillations started in November 2021 and were not present in earlier observations taken since the beginning of ZTF operations in 2018. 
\textit{Swift}/BAT data do not show indications of a previous oscillatory pattern, and neither \textit{Gaia} nor WISE data allow us to draw firm conclusions regarding previous variability
(see Appendix \ref{archivaldata} for details on these data).

To determine the nature of AT\,2021hdr, we first tested whether or not the variations can be explained by periodic or stochastic variations, and we find that they can hardly be explained in this way, although for the moment these models cannot be completely ruled out either (see Appendices \ref{period} and \ref{drw}).

Oscillations in the light curve of an AGN can be indicative of dynamic processes occurring near the SMBH. Several mechanisms can lead to such oscillations, and their detailed study provides valuable insights into the physics of accretion onto black holes. The specific timescales and characteristics of these oscillations can vary widely depending on the individual properties of the system, including both the mass and spin of the SMBH, and the nature of the surrounding environment. In the following, we show that the variations in AT\,2021hdr cannot be easily explained by any of the mechanisms usually associated to SMBHs. However, we find that the behavior of AT\,2021hdr broadly fits with models of the disruption and accretion of a gas cloud by a BSMBH \citep{goicovic2016}.

\subsection{Nuclear transients}

Transient events can result in strong continuum changes, sometimes with $\Delta$mag $> 1$. About 100 long-lasting strong flares have been detected in AGN showing large-amplitude variations not related to jet emission \citep{lawrence2016, graham2017, freerick2019}. However, the amplitude variations in AT\,2021hdr are too small to correspond to this type of flare ($\sim 0.2$ mag). 
Flaring events have also recently been observed in the X-rays as quasi-periodic eruptions (QPEs), where the X-ray count rate increases by up to two orders of magnitude within a timescale of hours, and these are recurrent events \citep[e.g.,][]{miniutti2019}. 
Current models for QPEs suggest that they result from an extreme mass-ratio inspiral (EMRI), where a companion with a mass comparable to that of the Sun collides with the accretion disk of a SMBH \citep{franchini2023, linialmetzger2023}.
However, all of them show a flat quiescent level, vary on timescales of hours, display very soft X-ray emission (below 2 keV), and have a spectral evolution that shows hysteresis in the relation between the bolometric luminosity and temperature \citep{arcodia2022, miniutti2023}. AT\,2021hdr shows no quiescence, a much harder X-ray spectrum, and its oscillations are seen from the X-rays to the optical.

More recently, monthly QPEs were claimed in X-rays from the nucleus of the nearby galaxy Swift J0230+28, which occur every $\sim$22 days \citep{evans2023, guolo2024}. The authors noted that this source shows spectral shapes and a temperature evolution that are distinct from those of the known QPE sources, probably indicating that the emission mechanism is likely quite different.
An alternative scenario to explain this source could include mass transfer from a star in an eccentric orbit around a SMBH, overflowing its Roche lobe each pericenter passage \citep{Krolik2022, Linial2023}.  
\cite{guolo2024} applied this model by considering a 33\rsun\ star in a proportionally wider orbit.  In order to match the observed period of AT\,2021hdr, an even larger star would be needed, and in any case the model does not account for the hard X-ray spectrum nor the UV/optical variations.  \cite{Metzger2022} proposed a related scenario, in which two stars orbiting around the SMBH in co-planar orbits have recurrent close encounters that result in tidal mass loss and subsequent accretion onto the SMBH.  This model can produce long-recurrence times, but the expected flares are very asymmetric, with a fast rise and slow decay given by the viscous timescale, unlike what we observed for AT\,2021hdr.

While accretion disks are usually assumed to be aligned to the black hole spin, their misalignment would cause the disk to precess, 
causing oscillations in the light curves as different parts of the disk are exposed to the observer \citep{liska2021, dotti2023}. In this scenario, for AT\,2021hdr to show oscillations at different wavelengths with the same apparent periodicity would require the disk to precess coherently all the way out to $R\sim0.01\,$pc, where the optical radiation is produced.  Even if that were possible, the precession period would be at least $\sim10^7\,$yr, many orders of magnitude longer than the observed variability period. Misalignment or wobbling of the inner accretion disk could 
also produce precessing jets that give rise to variations in flux \citep{liska2019}. The nondetection in VLBA observations of AT\,2021hdr casts doubt on this possibility, although we cannot exclude the presence of a nuclear compact jet detectable at higher frequency. More generally, the absence of modulations only two years ago necessitates an additional explanation for the turning on of the variability.

Some CL or changing-state AGN have shown strong changes in their light curves ($\Delta$ mag $> 0.5$) and in the profile of their broad emission lines. After these strong variations, however, CL AGN typically remain at these levels for years, while their variability remains stochastic throughout the event. 
AT\,2021hdr presents small-amplitude variability, and since the onset of the oscillations its variability has been in poor agreement with a stochastic process.
Furthermore, variations in the broad lines are not observed, as can be inferred from Fig. \ref{fig_halpha}, also ruling out a CL identification.

Another class of SMBH transients are caused by the tidal disruption of a star, namely TDEs, which are characterized by a rapid increase in the UV/optical with $\Delta$mag $> 2$, and sometimes declining as $t^{-5/3}$ \citep{vanvelzen2020, gezari2021}. 
Their spectra can show a large diversity but are generally blue and show weak [O III] lines, while at peak luminosity strong and broad HeII emission is present \citep{Charalampopoulos2022}. 
Other characteristics of TDEs include their constant color \citep{zabludoff2021} and their  usually very soft X-ray emission \citep{guolo2023}.

While most TDEs do not show periodic variations, \cite{Pasham2024} recently reported that AT2020ocn displays X-ray flares every 17 days, which these authors explain as the result of lense-thirring precession of a newly formed accretion disk. Crucially, this source does not
show similar variability in the optical--UV regime.  Regarding AT 2021hdr, the presence of oscillations at optical and UV would imply a larger-scale precessing accretion disk, as mentioned above. 
However, although the origin of the optical–UV emission from TDEs is still unknown, it is not usually associated to direct disk emission \citep[e.g.,][]{price2024}.
In addition, the relatively low Eddington ratio of AT\,2021hdr would imply that the inner disk can tear into discrete annuli that precess individually \citep{nixon2012} and consequently the X-ray modulations would be the result of a combination of changing orientation and accretion of discrete precessing annuli.
In this scenario, it would be unlikely to also observe a similar oscillation shape in the optical--UV from the outer part of the disk, and so we rule out this scenario for AT\,2021hdr.
Furthermore, theoretical studies have estimated that for a SMBH with a mass of 10$^7$M$\odot$, the disk will align on a timescale of about half a year, preventing further oscillations \citep{stone2012}, whereas the variations in AT\,2021hdr have been observed for almost 3 years and are ongoing.
Partial TDEs, which should also be recurrent sources, have been seen in only a few cases and their decline goes as $t^{-9/4}$ \citep{miles2020}. A popular example of a source that has been interpreted as a partial TDE is ASASSN-14ko, which shows a periodicity of $\sim 115$ days in its UV--optical light curves \citep{payne2021, payne2023}. Each of the outbursts in the light curve shows a ``fast-rise and slow-decay” pattern, and the optical spectra show a blue wing in H$\beta$ (but not in H$\alpha$) during the outburst that is not present in the quiescent state \citep{huang2023}.

AT\,2021hdr shows very different properties with respect to the TDEs described above: the amplitude of the variations is small ($\sim 0.2$ mag), it does not show any of the typical line features of a TDE in its optical spectrum (see Appendix \ref{optspectra}), it shows color variations (a ``bluer when brighter" behavior), its X-ray spectrum is rather hard, and the rise and fall times are very similar (see Sect. \ref{moonnitoring}). Furthermore, we fitted the post-2022 light curve with the model of decaying TDEs and compared the derived slope with the theoretical expectations, but could only find poor fits (see Appendix \ref{tdefitting} for details).

It is worth noting that most known TDEs have occurred in quiescent galaxies. AT\,2021hdr is located in an AGN, and it could therefore be that the phenomenon exhibits distinct characteristics. If a TDE occurs in an AGN, the interactions between the disrupted star and the SMBH can influence the properties of the AGN \citep{kathirgamaraju2017}. 
The simulations of \cite{chan2019, chan2020} actually show that this perturbation results in quasi-periodic behavior, but with a very short period associated to the dynamics of the inner disk, which does not fit AT\,2021hdr. Observationally, a few candidate TDEs have been reported in AGN \citep{ricci2020, zhang2023, Petrushevska2023}. 1ES 1927+654 could be the best case so far, and its overall properties are very different from those of AT\,2021hdr \citep{Trakhtenbrot2019ApJ, ricci2020, ricci2021, laha2022}.

\subsection{Binary SMBHs}

The orbital motion of a BSMBH is expected to result in strong periodic or quasi-periodic modulations of the mass-accretion rate and observed flux \citep[e.g.,][]{cuadra2009, lai2023}, and can cause shifting in the spectral lines \citep{popovic2012, dorazio2023}. We examine now whether a BSMBH system could explain the oscillations observed in AT\,2021hdr.

The timescales involved in the periodicities of observed BSMBH candidates are reported to be from a few hundred days to decades, implying sub-pc separations  \citep{valtonen2008, graham2015, charisi2016, chen2023}.
The BSMBH period itself will slowly decrease as the binary orbit shrinks due to gravitational wave emission \citep{peters1964}, although on much longer timescales (see below). For AT\,2021hdr, there may be periodicity in the flux, but the spectral lines have not changed. Additionally, AT\,2021hdr shows a well-defined starting point for the oscillations, which is not expected for a BSMBH.

TDEs on the other hand do have a well-defined start, and can happen in a BSMBH system \citep{amaro2013}. In this case, the binary affects the trajectory of the disrupted star, influencing the observational signatures of the TDE. Indeed, the first candidate for a TDE in a BSMBH was reported by \cite{liu2014}, who followed the method in \cite{liu2009} to find BSMBH in quiescent galaxies.

Numerical simulations of a TDE caused by a BSMBH \citep{coughlin2017, vigneron2018} have revealed that this interaction leads to a sudden increase in accretion followed by an overall decay with a power-law exponent of -5/3, but that interruptions to the accretion process occur due to the binary orbit. As a result, troughs appear superimpossed on the general TDE decaying light curve. These interruptions are periodic in some cases, with the period approximately corresponding to the binary orbital period.
This behavior is consistent with most of the observed evolution of  AT\,2021hdr,\footnote{Even though we do not have observations of the first peak 
in 2021 because the source was behind the Sun.}  
with the exception of the 2024 observations, which show a significant increase in flux (see Appendix \ref{tdefitting}).
Although we cannot be sure whether this flux increase is related to the oscillatory behavior or to intrinsic disk variability, it is clear that the overall shape and span shown in the light curves of \cite{vigneron2018} are inconsistent with observations of  AT\,2021hdr.
Indeed, a flux increase is not present in any of the models of \cite{vigneron2018}, which covered a large parameter space in terms of initial stellar orbit, binary mass ratio, and separation. However, those models did not consider the previous existence of an accretion disk, which should be the case for AT\,2021hdr given its Seyfert nature. To address whether the recent re-brightening could be due to the interaction between the TDE debris and a pre-existing disk requires a numerical effort that is beyond the scope of this work and will be addressed in a future publication when more data are available; however, we consider that a BSMBH TDE cannot be ruled out at this stage.

In addition to stars, gas clouds can also be disrupted by SMBHs and their binaries.  The key difference is that the clouds can be comparable to or even larger than the binary separation, unlike stars, which are always much smaller. \cite{goicovic2016} modeled the disruption and posterior evolution of gas clouds five times more extended than the BSMBHs they fall onto.  These authors tested different relative orientations and impact parameters, and measured the resulting accretion rate as a function of time, finding that the accretion rate has large peaks, separated by half the binary period, as each black hole crosses the cloud on its orbital period. Unlike the stellar case studied by \cite{vigneron2018}, however, different parameters result in diverse behaviors, including the number of peaks and the ratio between their maximum amplitudes.  Visual inspection of their results \citep[their Fig.~4]{goicovic2016} suggests that AT\,2021hdr could be produced by a cloud that approached on a trajectory perpendicular to the binary orbit, with an impact parameter larger than (but comparable to) the binary radius.  We acknowledge that, in reality, the parameter space of cloud properties (sizes, shapes, density distribution,  etc.) is enormous, and so it is likely that other configurations could also reproduce the observed behavior.

  \begin{figure}
   \centering
   \includegraphics[width=10cm]{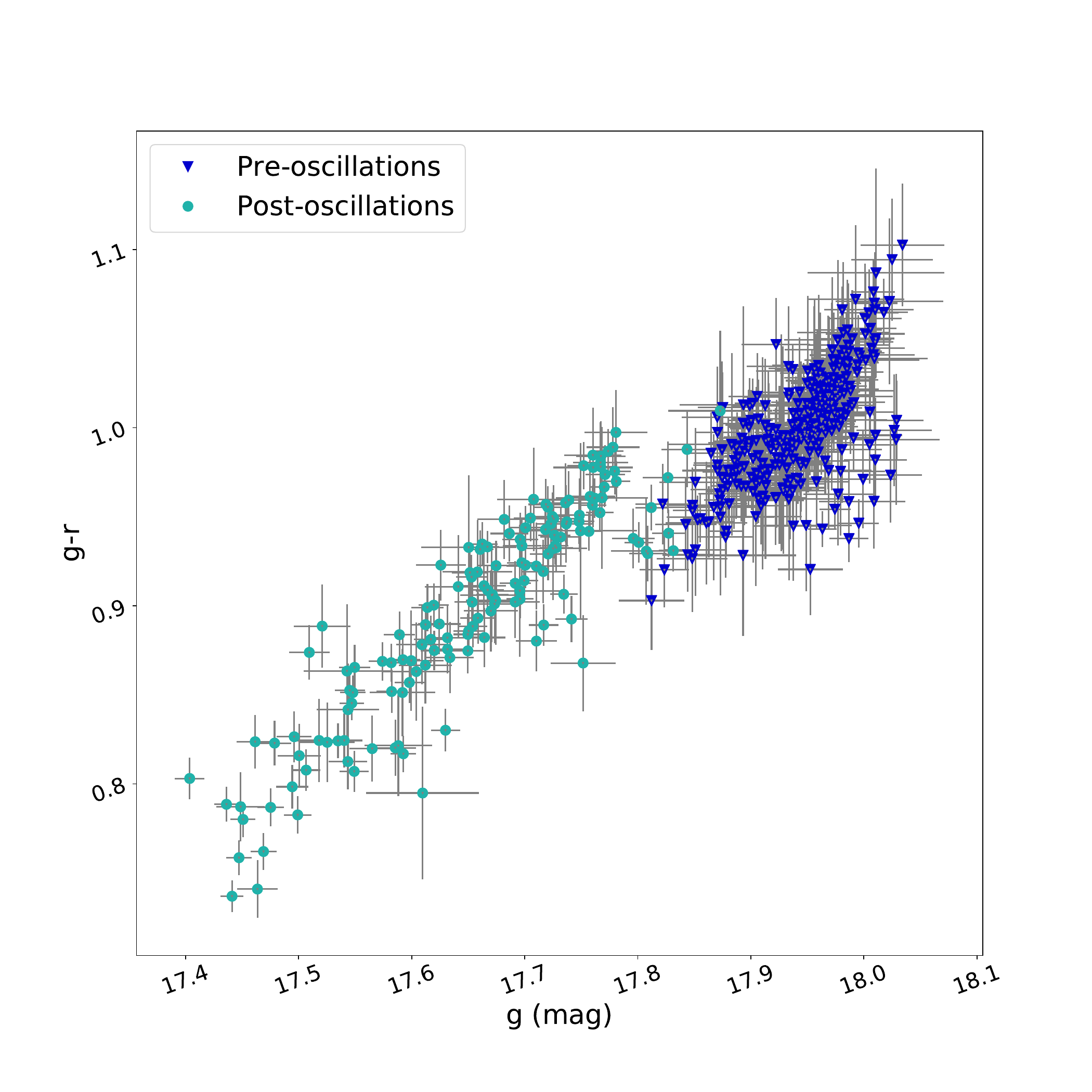}
      \caption{Color vs magnitude diagram. The color is obtained from the apparent magnitudes and the magnitude is in the g band. The blue triangles and cyan circles represent dates before and after the oscillations started in AT\,2021hdr (MJD = 59520). We note that this date does not correspond to the first alert but to the date when oscillations begin. }
         \label{fig_colmag}
   \end{figure}

In the scenario in which a cloud is responsible for the changes, we expect that the cloud-unbound debris could obscure and redden the central source. In Fig. \ref{fig_colmag} we plot the g-r color estimated from the ZTF apparent magnitudes against the g band magnitude. The blue triangles represent dates before the source started to show oscillations (MJD $<$ 59520), and cyan circles represent dates during the oscillatory pattern. A clear break exists before and after the oscillations occur, qualitatively agreeing with the cloud disruption model.

Following the \cite{goicovic2016} model, we can estimate the mass of the accreted cloud. We attribute the measured change in bolometric luminosity of $\Delta L \approx 1.7\times10^{44}\,$erg/s to accretion, using $\Delta L = \epsilon \dot M c^2$ ---with $\epsilon\approx0.06$ being the radiative efficiency of the accretion flow---, as  the initial, pre-oscillation Eddington ratio of $R_{\rm Edd}$ places the galaxy already in the standard accretion regime.  The typical accretion rate during the oscillations is therefore $\dot M \sim 0.05 \msun$/yr, and the accreted mass so far is $\Delta M \sim 0.1 \msun$. The relevant simulations of \cite{goicovic2016} show that a fraction of between 3 and 30 percent of the cloud is accreted after a few cycles, and so the initial mass of the cloud would be in the range of 0.3 to 3\msun.  As accretion is likely overestimated in the simulations given the large sizes of the sink particles modeling black holes, these values are likely a lower limit for the initial cloud mass.
Still, this kind of gas cloud could not possibly be directly observed outside our own Galaxy.  Observations of our Galactic center have revealed the presence of gas clouds and streams undergoing tidal disruption, but with mass estimates in the range of 3--50 \textit{Earth} masses only \citep[e.g.,][]{Gillessen2012,Ciurlo23}.  The required cloud in our model rather matches the cores observed a few hundred pc further out, inside the clouds of the central molecular zone (CMZ).  \cite{Lu2020} found almost 1000 such cores, with masses in the range of 0.3--300\msun\ and sizes of 5--40 mpc.  Theoretically, hydrodynamical simulations show that cloud formation and their infall to the central black hole(s) are a common feature in galactic nuclei given the effects of turbulence and self-gravity, and are able to efficiently drive AGN accretion \citep[e.g.,][]{Hobbs2011, Fiacconi2013}.

We can also describe the orbit and evolution of our putative binary, with a total mass of $4\times10^7\,\msun$ (see Appendix \ref{analysisoptical}), and an orbital period of $\approx 130\,$d, which is twice the observed peak recurrence time. 
This binary would have a separation of $\approx 0.83\,$mpc, which corresponds to $\approx 220 - 430\,R_{\rm Sch}$ for the primary black hole, depending on the binary mass ratio. 
This separation is much smaller than the expected size of the broad line region, which is consistent with the fact that AT\,2021hdr shows only one set of lines, and that they do not shift their wavelength with time. 
Additionally, this short separation means the binary is well inside the gravitational-wave-emission regime. Using the formula of \cite{peters1964}, and assuming a circular and equal-mass binary, we can estimate a timescale for orbital decay of $\approx 7\times 10^4\,$yr.  
Other processes, such as stellar scattering \citep[e.g.,][]{yu2002}, accretion from a circumbinary disk \citep[e.g.,][]{valli2024}, or the continuous infall of clouds \citep{goicovic2017}, will have a much slower effect on the orbital evolution.
The gravitational-wave-driven decay will be faster for eccentric binaries, but slower for binaries of unequal mass, and it is also very sensitive to the uncertain  estimate of the total black hole mass. In any case, a decay timescale of $\sim 10^5\,$yr, and the fact that there is another AGN $\sim 9\,$kpc away, make the source a very interesting target for studying the hierarchical growth of SMBHs.
We note that similar estimates can be carried out if the source is produced by a stellar TDE, the main difference being that the orbital period would then correspond to the observed period of $\approx 60-90\,$d, and so the binary would be more compact and decay faster.

\section{\label{summary}Summary}

We present AT\,2021hdr, a transient source discovered in the ZTF public alert stream, whose location coincides with the nucleus of a low-redshift Seyfert galaxy.
This galaxy is located at 9 kpc from a companion galaxy with which it is merging. This merger is reported in this work for the first time.
Since the initial alert, the source has shown an oscillatory pattern, and five  $\sim$0.2 mag  increases in brightness have been observed, roughly every $\sim$60-90 days. The same pattern is observed in the UV and X-rays with \textit{Swift}. Optical spectra were obtained at different dates and flux states, and showed that the spectral lines do not show changes in width or flux (see Appendix \ref{analysisoptical} and Fig. \ref{fig_halpha}).
VLASS images show radio emission, but sensitive high-resolution VLBA observations did not detect a compact source, ruling out a jet origin. 

While the oscillatory behavior suggests a BSMBH, archival data show that, before the alert, the source presented a roughly constant flux, and lower luminosity. A tidal disruption event, even one in a binary, does not fit the observations either, as the overall flux evolution does not follow the expected decay trend. We therefore propose that the source behavior could be due to the tidal disruption of a gas cloud by an unresolved BSMBH (not related to the companion galaxy at 9 kpc).

This process was modeled numerically by \citet{goicovic2016}, who found that the binary accretes the gas in several successive peaks (two per orbit), with the relative peak intensities depending on the orbital configuration. 
The initial mass of the accreted cloud would be in the range of 0.3-3\msun. 
Given an estimate of $\sim$4$\times$10$^7$ M$_{\odot}$ for the SMBH masses, we can calculate that the putative binary has a separation of $\sim$0.83 mpc and will merge in $\sim$7$\times$10$^4$ years, making it an interesting target for studying the hierarchical assembly of SMBHs in the local Universe.

Continued monitoring will allow us to constrain or rule out this scenario, and to inform the development of numerical models tailored to this interesting source.

\section*{Data availability}

The log of the ZTF forced photometry and \textit{Swift} observations are available and can be downloaded at Zenodo via \url{10.5281/zenodo.13942552}.

\begin{acknowledgements}
      We thank the referee for her/his comments and suggestions that helped to improve the paper. We warmly thank Felipe Goicovic for the new visualizations of his models that we use in the press release of this article.
      We acknowledge funding from ANID programs:  Millennium Science Initiative ICN12\_009 (LHG, AMMA, FEB, FF, MC), FONDECYT Iniciación 11241477 (LHG), CATA-BASAL FB210003 (FEB, CR), BASAL FB210005 (AMMA), FONDECYT Regular 1240875 (PL), 1200495 (FEB), 1230345 (CR), and 1231637 (MC), Millennium Science Initiative Program NCN$2023\_002$ (PL, PA, JC).

      JR and GB acknowledge financial support under the INTEGRAL ASI/INAF No. 2019-35.HH.0 and funding from the European Union’s Horizon 2020 Programme under the AHEAD2020 project (grant agreement n. 871158).

      IM, JM, MPS and ADM acknowledge financial support from the Severo Ochoa grant CEX2021-001131-S funded by MCIN/AEI/10.13039/501100011033, and the Spanish MCIU grant PID2022-140871NB-C21. MPS acknowledges financial support from the Spanish MCIU under the grant PRE2021-100265, which is part of the project SEV-2017-0709.

      This work was carried out with the help of the CONAHCYT research grants 280789 and 320987. Also, this work was supported by the MPIfR-Mexico
 Max Planck Partner Group led by V.M.P.-A. A.G.-P. acknowledges support from the CONAHCyT program for her PhD studies.

The ASKAP radio telescope is part of the Australia Telescope National Facility which is managed by Australia’s national science agency, CSIRO. Operation of ASKAP is funded by the Australian Government with support from the National Collaborative Research Infrastructure Strategy. ASKAP uses the resources of the Pawsey Supercomputing Research Centre. Establishment of ASKAP, the Murchison Radio-astronomy Observatory and the Pawsey Supercomputing Research Centre are initiatives of the Australian Government, with support from the Government of Western Australia and the Science and Industry Endowment Fund. We acknowledge the Wajarri Yamatji people as the traditional owners of the Observatory site. This paper includes archived data obtained through the CSIRO ASKAP Science Data Archive, CASDA (https://data.csiro.au).
The National Radio Astronomy Observatory is a facility of the National Science Foundation operated under cooperative agreement by Associated Universities, Inc. 

      We acknowledge the use of public data from the \textit{Swift} data archive through ToO proposal number IDs,

      18127, 
      18147, 
      18601, 
      18817, 
      18823, 
      19047, 
      19202, 
      19317, 
      19522, 
      19751, 
      19967, 
      18127, 
      18147, 
      18601, 
      18817, 
      18823, 
      19047, 
      19202, 
      19317, 
      19522, 
      19751, 
      19967, 
       and GI program \#2023221.
      
      The ZTF forced-photometry service was funded under the Heising-Simons Foundation grant
\#12540303 (PI: Graham).

    Part of this work was carried out using the services from the ALeRCE broker, in particular using its Web Interface and the ZTF Forced Photometry Notebook.

    The data presented here were obtained in part with ALFOSC, which is provided by the Instituto de Astrofisica de Andalucia (IAA) under a joint agreement with the University of Copenhagen and NOT.

    Based upon observations carried out at the Observatorio Astronómico Nacional on the Sierra San Pedro Mártir (OAN-SPM), Baja California, México.

    This work has made use of data from the European Space Agency (ESA) mission {\it Gaia} (\url{https://www.cosmos.esa.int/gaia}), processed by the {\it Gaia} Data Processing and Analysis Consortium (DPAC, \url{https://www.cosmos.esa.int/web/gaia/dpac/consortium}). Funding for the DPAC has been provided by national institutions, in particular the institutions participating in the {\it Gaia} Multilateral Agreement.
    
    This publication makes use of data products from the Wide-field Infrared Survey Explorer, which is a joint project of the University of California, Los Angeles, and the Jet Propulsion Laboratory/California Institute of Technology, funded by the National Aeronautics and Space Administration.
    
    This publication makes use of data products from the Near-Earth Object Wide-field Infrared Survey Explorer (NEOWISE), which is a joint project of the Jet Propulsion Laboratory/California Institute of Technology and the University of Arizona. NEOWISE is funded by the National Aeronautics and Space Administration.
\end{acknowledgements}

\bibliographystyle{aa}
\bibliography{bibliography} 

\appendix

\section{\label{archivaldata}Archival data}

We gathered information from different instrumentation with observations before ZTF observed to check for previous variability. Here we report on observations with \textit{Gaia}, WISE and \textit{Swift}/BAT.

\subsection{\label{swiftbat}\textit{Swift}/BAT 157 month}

The Burst Alert Telescope (BAT) onboard the \textit{Neil Gehrels Swift }observatory has a main
objective that is to detect transient gamma-ray bursts (GRBs). To fulfill this objective it is continuously performing an all-sky hard X-ray survey in the 14-195 keV energy band \citep{tueller2008, tueller2010}. The last catalog is the 157-month survey\footnote{https://swift.gsfc.nasa.gov/results/bs157mon}, which contains light curves and spectra of $>$1800 sources (Lien et al. in prep).

\begin{figure}
   \centering
   \includegraphics[width=9.5cm,scale=0.5]{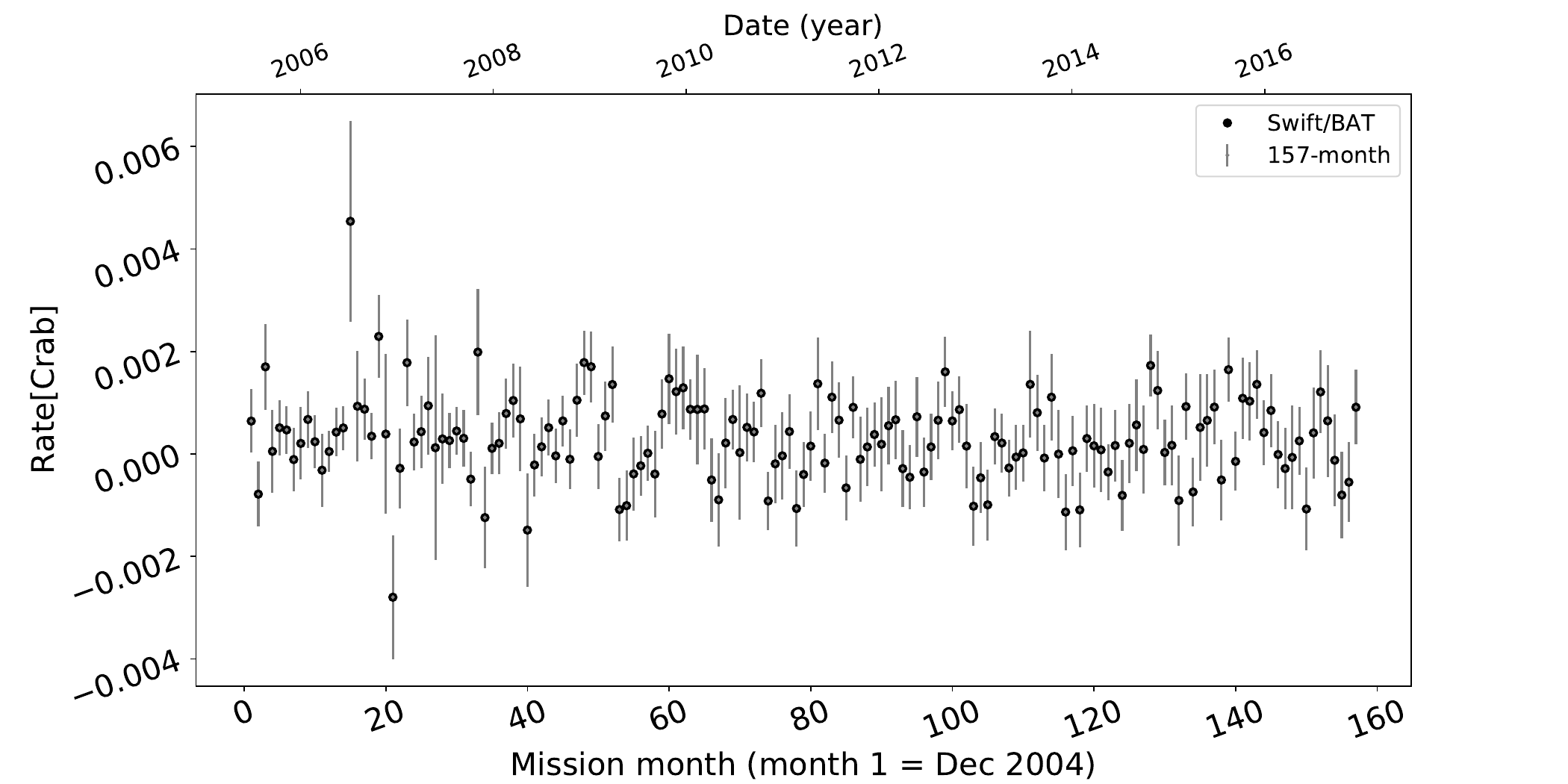}
      \caption{\textit{Swift}/BAT 157-month light curve of SWIFT J2123.9+3401 in the 14-195 keV band.
              }
         \label{fig_bat}
   \end{figure}

In Fig. \ref{fig_bat} we present the light curve of SWIFT J2123.9+3401 between 2004 and 2017, which is located at RA=320.982\textdegree, Dec=34.018\textdegree. The count rate is normalized using the Crab nebula as a standard candle \citep{oh2018}.

The source has a mean flux of 7.04$^{+2.05}_{-2.45}$ $\times$ 10$^{-12}$erg cm$^{-2}$s$^{-1}$, corresponding to a luminosity of logL(14-195 keV)= 44.0 erg s$^{-1}$.

We evaluated if the observed changes in the light curve are significant. The mean value of the count rate is 3$\pm$8$\times$10$^{-4}$ cts s$^{-1}$. We calculated the $\chi^2$ and degrees of freedom and obtained 165.2/156, indicating that this is a good fit respect to a constant value. Following \cite{vaughan2003} we estimated the normalised excess variance, $\sigma_{NXS}$=1.7$\pm$0.9 and the fractional variability F$_{var}$=1.2$\pm$0.4. These measurements indicate changes at a 1.9$\sigma$ and 2.7$\sigma$ of confidence level, therefore we will consider that the source did not vary in the 14-195 keV band between 2004-17.

\subsection{Gaia}

AT 2021hdr has a counterpart within 1'' in both \textit{Gaia} DR2 \citep{GaiaCollab2018Brown} and DR3 \citep{GaiaCollab2023Vallenari} catalogs, with source\_id = 1855138731842796032. In \textit{Gaia} DR3, it has mean magnitudes of $18.82 \pm 0.02$, $17.74 \pm 0.03$, and $16.22 \pm 0.01$ mag in G, BP, and RP filters, respectively. It has BP-RP color of 1.51 mag and a BP/RP excess factor of 6.54 mag; together they give a corrected BP/RP excess factor of 5.27 mag \citep{Riello2021}, indicating a consistency issue between \textit{Gaia} fluxes.

This object has phot\_variable\_flag = 'not\_available' in both data releases, indicating that it was not processed and/or exported. However, \citet{Mowlavi2021} estimated variability amplitude proxies for this object of $\approx$ 0.13, 0.12, and 0.04 mag for G, BP, and RP filters, respectively, based on \textit{Gaia} DR2 data.

This \textit{Gaia} source also appears in the \textit{Gaia} Focused Product Release catalog as a possible gravitational lens candidate \citep{GaiaCollab2023Krone-Martins}, including three components within 1''. Raw photometry light curves for these components are publicly available; however, as they are based on uncalibrated onboard magnitudes, we exclude these measurements from our analysis of AT 2021hdr variability.

\subsection{Wide-field Infrared Survey Explorer (WISE)}

\begin{figure}
   \centering
   \includegraphics[width=9cm]{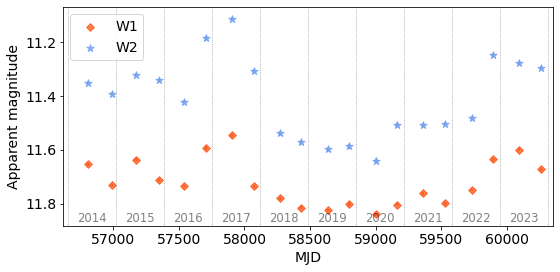}
   \includegraphics[width=9cm]{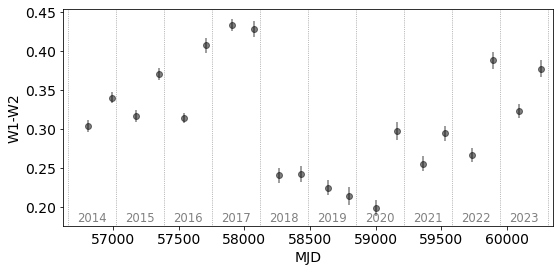}
    \includegraphics[width=9cm]{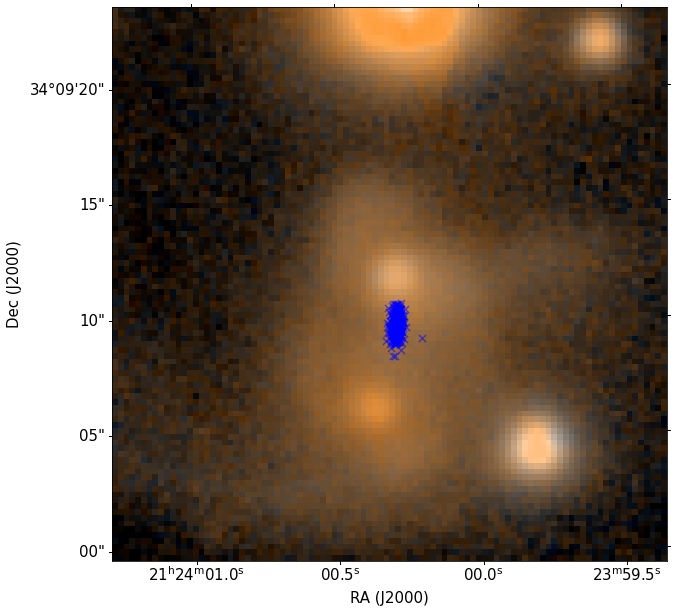}
      \caption{NEOWISE-R light curves (top panel) and color evolution (middle panel) based on WISE detections within 5'' of AT 2021hdr. The bottom panel is a PansTARRS image with the NEOWISE-R positions where the photometry was done. It can be seen that it falls in the middle of the two galaxies.}
         \label{fig_wise_lc}
   \end{figure}

We searched for counterparts to AT 2021hdr in the Wide-field Infrared Survey Explorer (WISE; \citealt{Wright2010}) data, starting with AllWISE Data Release \citep{Cutri2013}. This includes objects detected on the deep AllWISE Atlas Intensity Images, which are coadds based on WISE cryogenic and the Near Earth Object Wide-field Infrared Survey Explorer (NEOWISE; \citealt{Mainzer2011}) post-cryogenic survey phases. AT 2021hdr has a counterpart within 2'' in AllWISE, named J212400.31+340910.2 with unique source ID 3216134801351011461. This object has high ($> 20$) S/N detections in all WISE bands, with Vega magnitudes of $11.48 \pm 0.02$, $11.07 \pm 0.20$, $8.33 \pm 0.02$, and $6.01 \pm 0.05$ mag in the W1, W2, W3, and W4 filters. It has a variability flag $\lesssim 5$ in all WISE bands, indicating this object is most likely not variable. However, the W1 measurement is flagged as spurious or contaminated by the scattered light halo that surrounds a nearby bright source. The second closest AllWISE object to AT 2021hdr lies at 12''.

We also searched in the NEOWISE Reactivation mission (NEOWISE-R; \citealt{Mainzer2014}) latest Data Release, which covers the first ten years of survey operations. We queried all sources within 5'' of AT 2021hdr in the NEOWISE Single-exposure Source Database, which includes all detections that have combined W1 and W2 S/N > 3. We selected detections from good quality framesets by applying the following criteria: overall frameset quality score qual\_frame > 0; cc\_flags = 0000 (i.e. no filters were flagged as spurious detections or real sources contaminated by image artifacts); and profile-fit magnitudes 11 < W1 < 15 and 10 < W2 < 13 (i.e. bright enough detections but minimizing spurious transient detections like charged particle strikes, satellite trails and hot or noisy pixels). This gives 324 detections that lie within 1.2'' and 3.5'' of AT 2021hdr, with individual coordinate uncertainties $\lesssim 0.03''$. All of them are found in the region between the host and neighbor galaxy nuclei, and come associated to the same AllWISE object mentioned above.

We compute weighted means and weighted mean errors using the multiple magnitude measurements per observation epoch. This results in two magnitude estimates per filter per year, which we use to obtain color estimates as shown in Fig. \ref{fig_wise_lc}. The amplitude of the magnitude variation reaches 0.3 and 0.5 mag in W1 and W2 respectively, while for the W1-W2 color it reaches 0.25 mag. Binned magnitudes show a trend of increasing brightness from the first half of 2021 (coincident with the first ZTF alert date) to the first half of 2023, hinting a 6-month delay between the brightest ZTF peak and the brightest subsequent NEOWISE-R magnitudes. However, the fact that this emission is distributed in between the AGN system prevent us to fully associate this MIR emission to the nuclear transient (see bottom panel in Fig. \ref{fig_wise_lc}).

\section{\label{analysisoptical}Analysis of optical spectra}

\begin{figure}
   \centering
   \includegraphics[width=9cm]{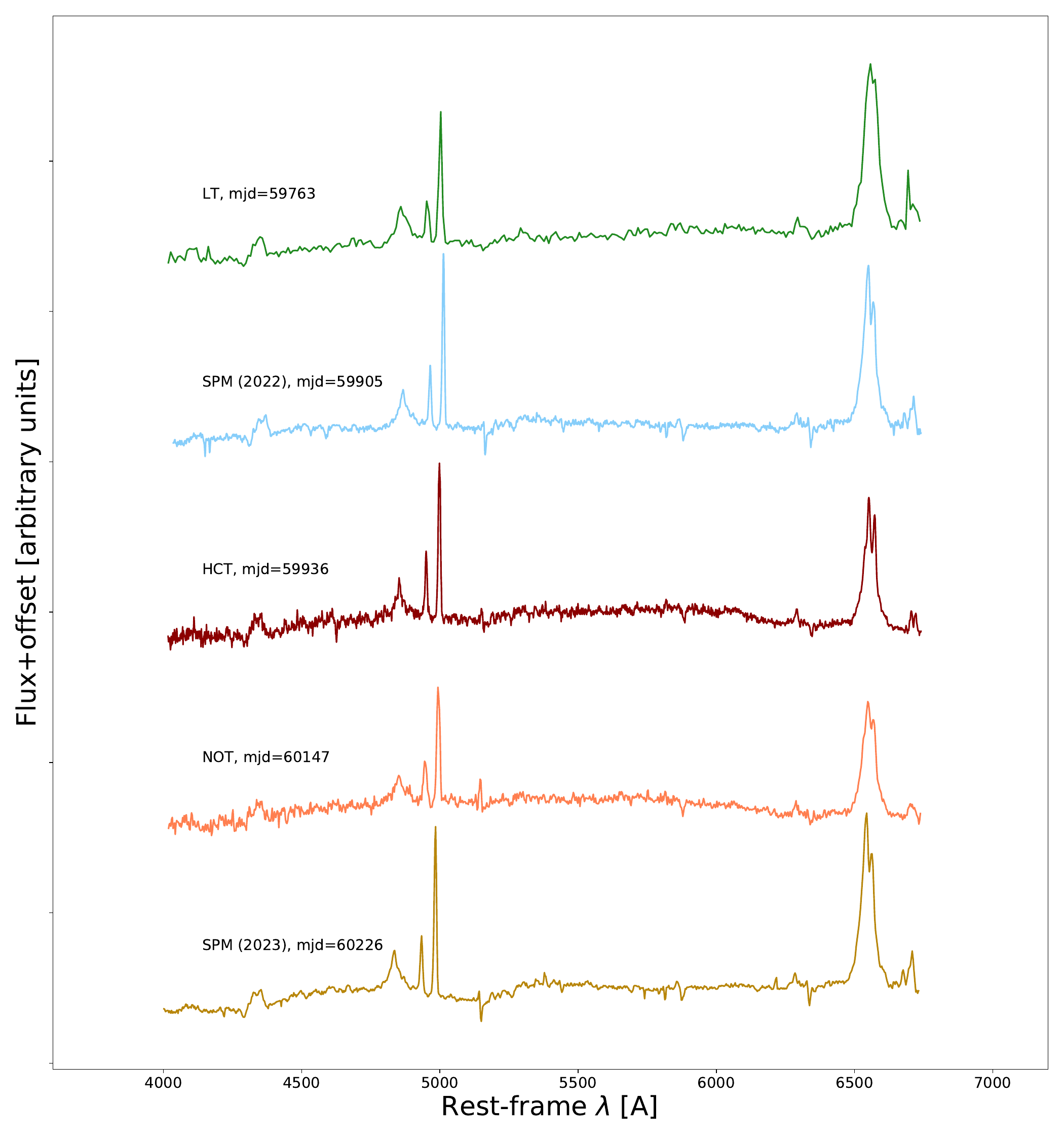}
      \caption{Optical spectra of AT\,2021hdr used in this work. From top to bottom, spectra from the LT, SPM, HCT, NOT, and SPM are presented in in chronological order. The y-axis units are arbitrary and the spectra are plotted with an offset for visualization purposes.}
         \label{fig_opt_spectra}
   \end{figure}

In this section we present details on the optical spectra introduced in Sect. \ref{optspectra}. In Fig. \ref{fig_opt_spectra} we present the spectra from the LT, SPM (2022), HCT, NOT, and SPM (2023) in chronological order and arbitrary units.

We will compare the two spectra taken with the SPM telescope because these were taken with the same setup and thus they are the most likely to be able to find spectral changes. The spectrum from 2022 was taken at the bottom of an oscillation, and the spectrum from 2023 at the top of an oscillation.

We used the Penalized Pixel-Fitting (pPXF) software \citep{cappellari2017improving} to fit the data, and the E-MILES library \citep{vazdekis2010evolutionary} to account for the stellar continuum component. The other components used to fit the AGN continuum and the  emission lines are:

\begin{itemize}
\item A power law template for the accretion disk contribution of the form $(\frac{\lambda}{\lambda_N})^\alpha$ where $\lambda$ is the wavelength, $\lambda_N= 5000$ \AA\ is a normalization factor and $\alpha$ goes from $-$5 to $-$0.1 in steps of 0.1.
\item Two components with permitted and forbidden emission lines, with free normalizations, to model the narrow lines. We allowed H$\alpha$ and H$\beta$ to have different velocity dispersion by considering the separation of the templates in $\lambda=6200$ \AA.
\item  Two components with permitted emission lines, with free normalizations, to model the broad emission lines. Using the same value of $\lambda=6200$ \AA, we allowed H$\alpha$ and H$\beta$ to have different kinematic moments (velocity and velocity dispersion).
\end{itemize}

We obtained errors for each parameter by performing Monte Carlo simulations using the best-fitting model and simulating random noise generated from the standard deviation of the best-fitting residuals. 

We performed a relative calibration between the spectra using data from the ZTF photometry. The procedure was to integrate the flux of the spectra in the g and r band and normalize the ZTF light curve and this measurement in the same date. Finally we use the ratio of these measurements to have a proper relative calibration between the spectra and the light curve.

\begin{figure}
   \centering
   \includegraphics[width=10cm]{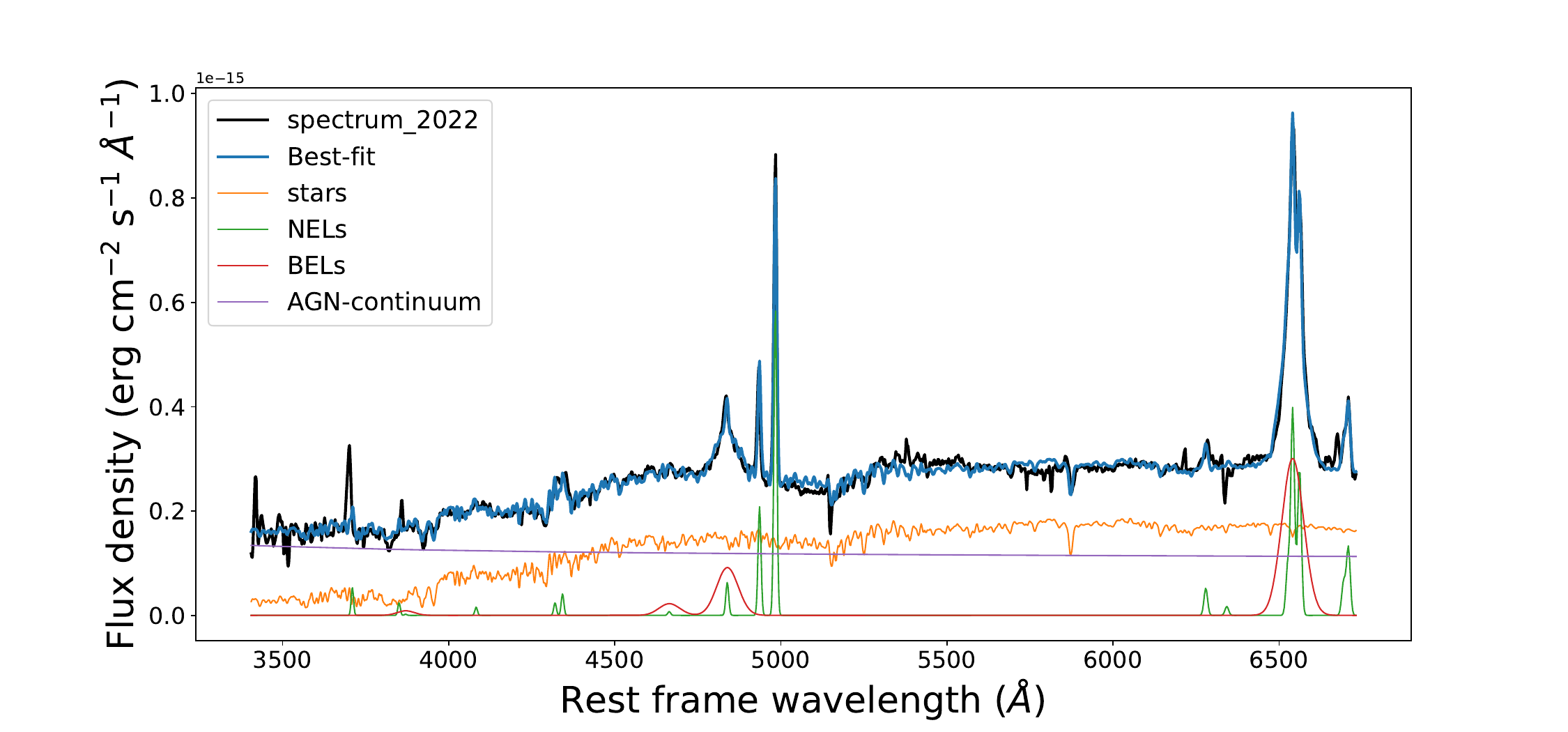}
   \includegraphics[width=10cm]{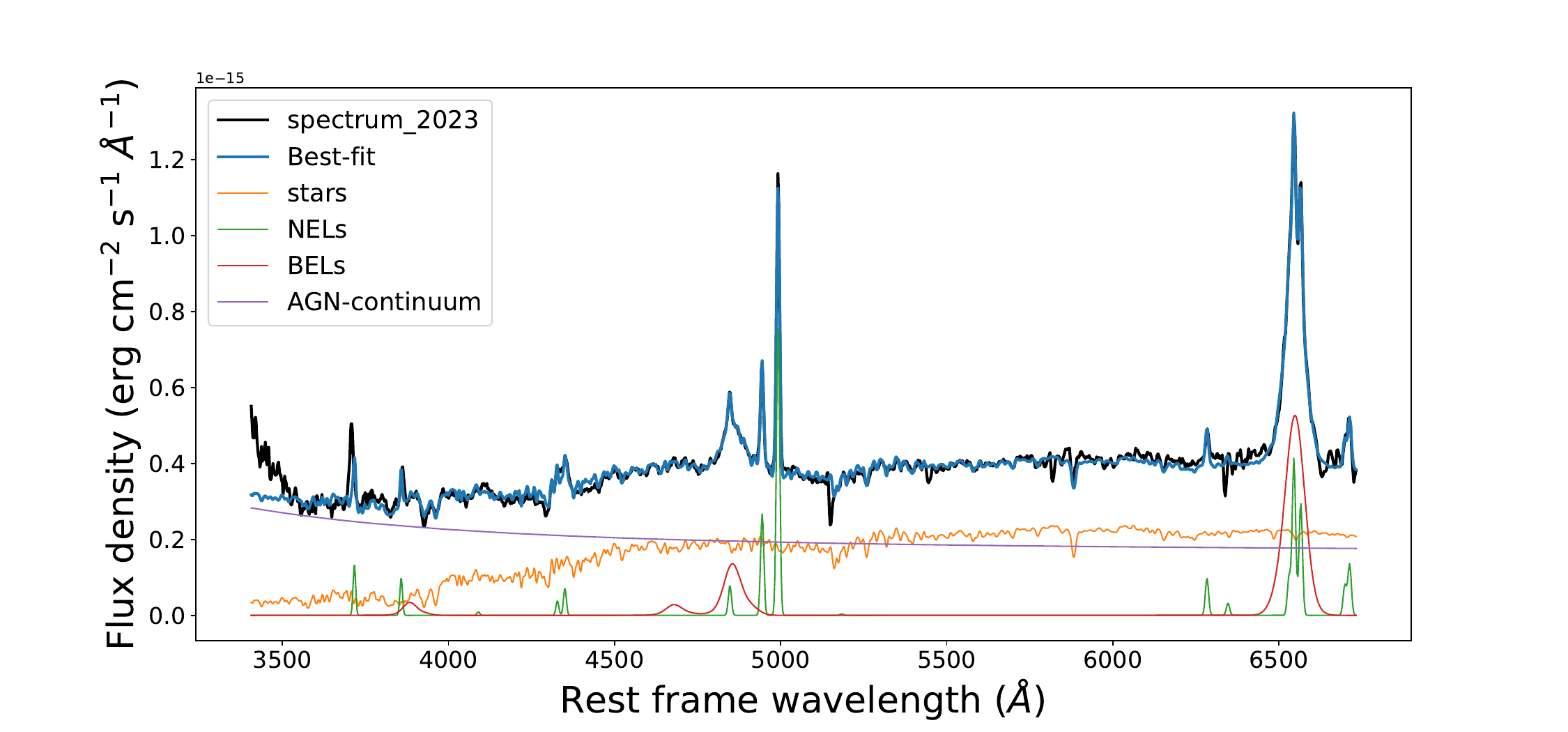}
      \caption{Decomposition of the optical spectra of AT\,2021hdr taken with the SPM in 2022 (top) and 2023 (bottom). The black corresponds to the original spectrum, the blue to the best-fitted model, the orange to the host galaxy contribution, the purple the continuum from the AGN emission, the green to the narrow emission lines (NELs),
      and the red to the broad emission lines (BELs).}
          \label{fig_fit_optspec}
   \end{figure}

 The decomposed spectra are presented in Fig. \ref{fig_fit_optspec}, where the best fit to the spectra is shown in blue.

  Table~\ref{tab: spectra} shows the results of the spectral fitting for the SPM spectra. The broad H$\alpha$ and H$\beta$ fluxes show negligible variation between the two spectra. However, when comparing the AGN-continuum luminosity at 5100\AA\ and 3500\AA\ we observed a noticeable variation in this component. This change is evident in the plots of Fig. \ref{fig_fit_optspec}, where the AGN component shows a harder emission in the SPM 2023 spectrum. Because our model for the AGN continuum is a combination of power laws and the profiles of both the 2022 and 2023 spectra, particularly for short wavelengths, we cannot directly compare the slopes. Instead, for a fairer comparison, we note that the contribution of the most negative slope ($\alpha = -5$, from the model $(\frac{\lambda}{\lambda_N})^\alpha$) is 1.7\% for the 2022 observation and 7.5\% for the 2023 observation.

  Narrow line fluxes give ratios log([O III]\,$\lambda5007$/\,H$\beta$) = 1.1, log([N II]\,$\lambda6583$/\,H$\alpha$) =  -0.01, and log([S II]\,$\lambda(6716{+}6731)$/\,H$\alpha$) = -0.3 (completely consistent between 2022 and 2023). These values clearly confirm the Seyfert nature of AT2021hdr. 

\begin{table*}[]
    \centering
    \caption{Results from the optical spectral fitting of the SPM spectra.   }
    \begin{tabular}{l|cccccc}
    \hline
        Spectrum     &  Flux H$\alpha$    & FWHM H$\alpha$      & Flux H$\beta$  & \multicolumn{2}{c}{AGN-Luminosity} & logM$_{BH}$ \\
            & ($10^{-14}$ erg cm$^{-2}$ s$^{-1}$) & (km s$^{-1}$)&($10^{-14}$ erg cm$^{-2}$ s$^{-1}$) &  \multicolumn{2}{c}{($10^{43}$ erg s$^{-1}$)} & (M$_{\odot}$) \\
             & & & & (5100\AA) & (3500\AA)\\
    \hline
        SPM-2022  &  3.49 $\pm$ 0.13 & 3525.00 $\pm$ 0.01 &1.05 $\pm$ 0.03 &1.31 $\pm$ 0.01 & 1.47 $\pm$ 0.01
        & 7.61 $\pm$ 0.50 \\
        SPM-2023 &  4.04  $\pm$ 0.14 & 3290.00 $\pm$ 99.19 & 1.01 $\pm$ 0.04& 1.51 $\pm$ 0.06 &2.133 $\pm$ 0.002 & 7.57 $\pm$ 0.51 \\
    \hline
    \end{tabular}
    \tablefoot{Flux and FWHM columns are for broad emission lines. In the luminosity columns we compare the AGN continuum between the two spectra at two different wavelengths. For the calculation of the luminosity we used a $\Lambda$CDM cosmological model with $H_0=70$ km/s/Mpc. The black hole masses are derived from the virial mass estimator using H$\alpha$.   
    We added uncertainties of 0.5 dex to the error of the masses, as expected for virial mass estimates 
    \citep{reines2015}.  }
    \label{tab: spectra}
\end{table*}

We estimated the black hole mass from these spectra using the single-epoch virial mass estimator \citep[e.g.,][]{reines2015}. From both spectra we obtained a mass estimation of M$_{BH} \sim$ 4$\times$10$^7$ M$_{\odot}$. The luminosity was obtained 
with a $\Lambda$CDM cosmology with parameters  $\Omega_M=0.3$, $\Omega_{\lambda}=0.7$ and $H_0=70$ kms$^{-1}$ Mpc$^{-1}$.

\begin{figure}
   \centering
   \includegraphics[width=9cm]{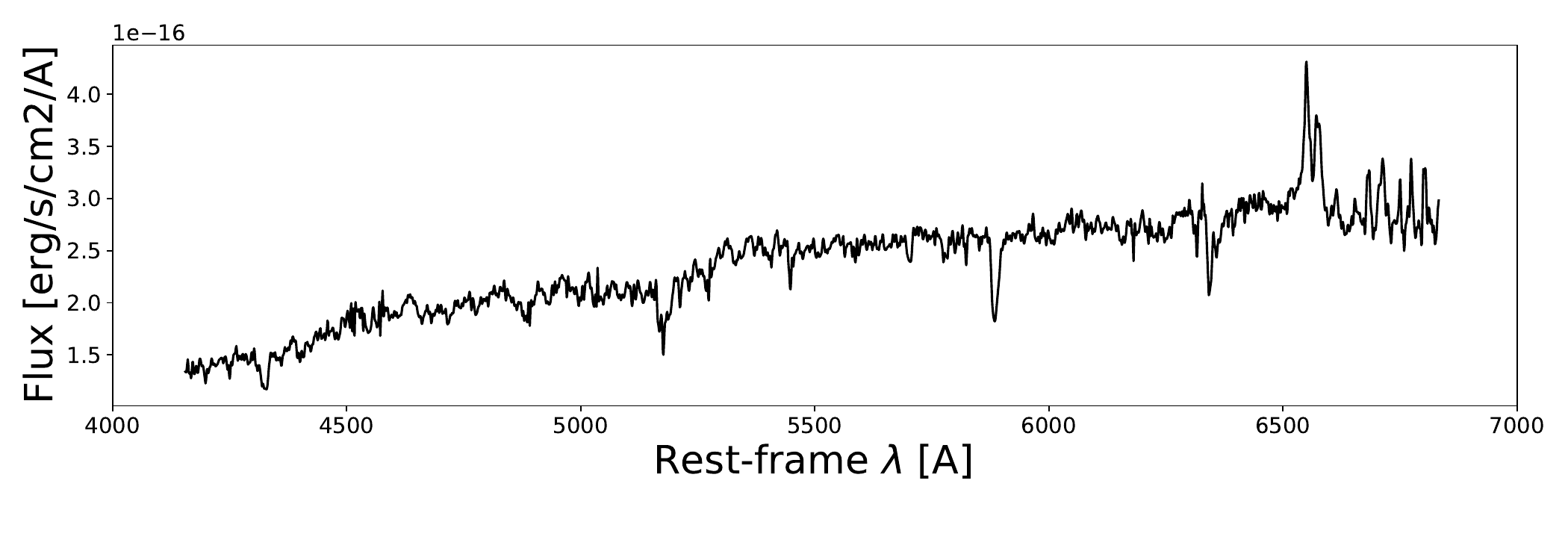}
      \caption{Optical spectrum of the neighbour galaxy located 6 arcsec (9 kpc) away from the nucleus of AT\,2021hdr taken with the SPM in 2022.}
         \label{fig_opt_spectra_neightbour}
   \end{figure}

In Fig. \ref{fig_opt_spectra_neightbour} we present an optical spectrum taken with the SPM of the neighbour galaxy,  2MASS J21240037+3409058, located at 6" or $\sim 9$ kpc to the south. The redshift of the galaxy is 0.081. This is the first spectrum reported for this source, allowing its classification as a LINER nucleus.

\section{Possible origin of the radio emission\label{radioemission}}

The radio power, as estimated from the radio emission detected in VLASS and co-located with AT\,2021hdr, is $10^{38.4}$ erg s$^{-1}$, ascribing the source to the radio-quiet regime. However, the upper limit on the VLBA luminosity is  $10^{37.5}$ erg s$^{-1}$, implying that part of the arcsecond scale flux is resolved out at milli-arcsecond scales. Both fluxes are consistent with the range of radio powers found for the faintest Seyfert nuclei observed to date ($10^{35}-10^{39}$ erg s$^{-1}$, \citealt{2019MNRAS.485.3185C, pangiroletti2013}). The observed ratio between the radio luminosity (VLASS and VLBA) and the observed Swift X-ray luminosity is also consistent to that found for local Seyfert galaxies \citep{pan07, pangiroletti2013}. Possible origins for the radio emission include a low luminosity radio core, possibly present below the VLBA sensitivity. Higher frequency and sensitive observations would be needed to detect such possible core \citep{pan19}. Alternatively, diffuse radio emission coming from star-forming activity in the host bulge may be present. This would show a steep spectral index ($\sim$-0.7), as a result of synchrotron emission from supernovae ejecta, with a typical lifetime of 100 Myr \citep{1992ARA&A..30..575C}. It is also possible that recent ($\sim$10 Myr) star-forming activity could result in a flat radio spectrum from bremsstrahlung emission (\citealt{2017ApJ...836..185T}, and references therein), but still detectable only by integrating the emission on the kpc-scale. A star-forming origin would imply that there is no physical connection between the oscillating nuclear transient (pc-scale) and the radio emission (kpc-scale), which it could be related with the ongoing merger of the two galaxies, triggering star formation via gas compression. Eventually, wind shocks may also produce radio emission. Indeed, the radio luminosity was found to be correlated with wind velocities as estimated from the [O III] emission line in low-redshift ($z<0.8$) type 2 radio-quiet quasars, in the range $10^{38.5}-10^{41}$ erg s$^{-1}$ \citep{2014MNRAS.442..784Z}. This scenario can be further tested via intermediate resolution (sub-kpc) radio observations, combined with spectroscopic ones.


\section{Time series analysis \label{timeseries}}

Here we will use different methods to fit the ZTF light curves of AT\,2021hdr to evaluate if these variations could be explained by processes related to the AGN accretion disk, periodic movements or external sources such as TDEs. The results are discussed in Sect. \ref{disc}.

\subsection{Periodogram \label{period}}

We used P4J\footnote{https://github.com/phuijse/P4J} to calculate the periodogram of AT\,2021hdr. P4J is a python package for period detection on irregularly sampled and heteroscedastic time series based on Information Theoretic objective functions. The core of the used package is a class called periodogram that sweeps an array of periods/frequencies looking for the one that maximizes a given criterion. The main contribution of this work is a criterion for period detection based on the maximization of Cauchy-Schwarz Quadratic Mutual Information \citep{huijse2018}. Information theoretic criteria incorporate information on the whole probability density function of the process and are more robust than classical second-order statistics based criteria \citep{principe2010}. 

We obtain the multiband period from the implementation in the P4J library using the Multi Harmonic Analysis of Variance (MHAOV) periodogram \citep{mondrik2015}, using both g and r bands at the same time.

\begin{figure}
   \centering
   \includegraphics[width=10cm]{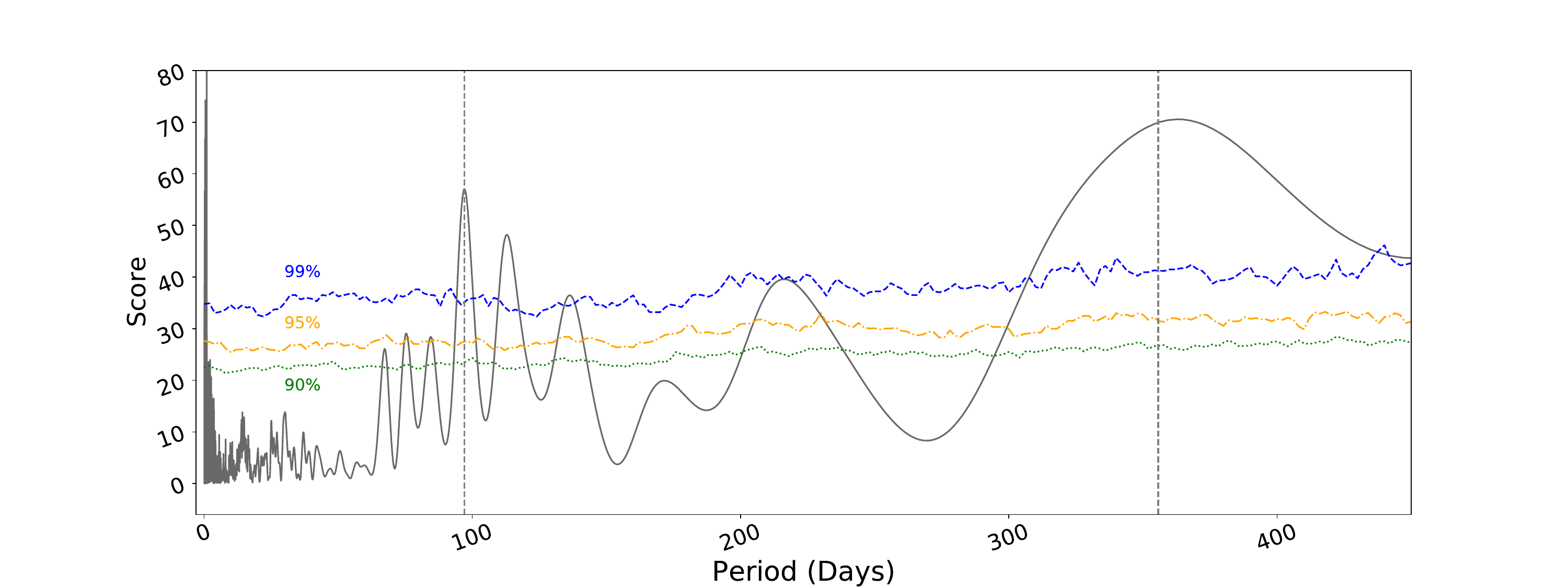}
      \caption{Periodogram of AT\,2021hdr using the P4J code using the ZTF g and r bands. The dashed lines represent the best periods. The horizontal dashed/dotted lines represent the 90, 95, and 99\% confidence levels.}
         \label{fig_periodogram}
   \end{figure}

Figure \ref{fig_periodogram} shows the resulting periodogram for the ZTF light curves for dates > 59500. The dashed lines represent the best periods, which are estimated as the local maxima, being 97.7, and 355.7 days. 
We estimated the confidence levels at 90, 95, and 99\% using 50000 iterations of Montecarlo simulations of the light curves
following the procedure in \cite{amaya2022}. The results are plotted as dashed lines. 
The obtained period could be significant, but because only a few oscillations have been observed so far, we prefer to wait for a longer light curve to better estimate its periodicity, and quasi-periodicity cannot be ruled out at this moment either. 

\subsection{Continuous-time autoregressive moving average (CARMA) models \label{drw}}

To test if the variations observed in AT\,2021hdr can be explained by a stochastic process, we used CARMA models to fit its ZTF light curve, in particular we tested a Damped Random Walk (DRW or CAR(1)) model  \citep[e.g.,][]{kelly2009, sanchezsaez2021} and a Damped Harmonic Oscillator (DHO or CARMA(2, 1)) model \citep{kelly2014, moreno2019}.
.

We use the EzTao Python package \citep{Yu2022}, that performs time-series analysis using CARMA processes. It uses celerite (a fast gaussian processes regression library) to compute the likelihood of a set of proposed CARMA parameters given the input time series. 
We fitted a DRW model, which has two process parameters of intrinsic characteristic variability amplitude, $\sigma_{DRW}$, and timescale, $\tau_{DRW}$.

Some authors argued that DRW models do not always
adequately describe the variability properties of AGN but that the slightly more complex DHO models apparently
may do so, as they can naturally produce the low-frequency break in the power spectral density \citep{kelly2014}. This model has four parameters of the amplitude, $a_1$ and $a_2$, and two for the timescale, $b_1$ and $b_2$:

\begin{equation}
    d^2x + a_1d^1x + a_2x = b_1 \epsilon(t) + b_2d( \epsilon (t))
\end{equation}

We run Markov Chain Monte Carlo (MCMC) to determine a distribution over the parameter space. The results of the fitting are presented in Table \ref{table:drw}. The errors of these parameters are too large, indicating that stochastic variations does not fit the ZTF light curve properly.

We also estimated the structure function following \cite{sanchezsaez2018}, using the Bayesian definition presented in \cite{Schmidt10}. We obtained a value of $0.30\pm 0.01$ for the logarithmic gradient of the variation in magnitude. This value is compatible with results obtained for type 1 AGN (e.g., \citealt{sanchezsaez2018,DeCicco22}), so from the structure function we cannot discard stochastic AGN-like variations.

%
\begin{table*}
\caption{Parameters for the DRW and DHO.}             
\label{table:drw}      
\centering                          
\begin{tabular}{ccccc}        
\hline\hline                 

  &   \multicolumn{2}{ c }{Alerts} & \multicolumn{2}{ c }{Total} \\
  & g & r & g & r \\
\hline
\multicolumn{5}{ c }{DRW} \\ \hline
$\tau_{DRW}$ & 0.15$_{-0.04}^{+0.13}$ & 0.10$_{-0.03}^{+0.10}$
 & 0.22$_{-0.07}^{+0.29}$ & 0.12$_{-0.02}^{+0.03}$ \\
$\sigma_{DRW}$ & 195$_{-88}^{+513}$ & 256$_{-123}^{+774}$ &  621$_{-332}^{+2652}$ & 105$_{-29}^{+63}$  \\
\hline
\multicolumn{5}{ c }{DHO} \\ \hline
$a_1$ & 0.6$_{-0.3}^{+6 \times 10^{40}}$  & 
5$\times 10^{67}$ $_{-5 \times 10^{67}}^{+1 \times 10^{119}}$ &  5$\times 10^{59}$ $_{-5 \times 10^{59}}^{+3 \times 10^{117}}$ & 193$_{-51}^{+67}$ \\
$a_2$ &  0.004$_{0.003}^{3 \times 10^{38}}$  &  
2$\times 10^{65}$ $_{1 \times 10^{65}}^{4 \times 10^{116}}$
&   5$\times 10^{56}$ $_{5 \times 10^{56}}^{3 \times 10^{114}}$ & 0.11$_{-0.09}^{+0.21}$ \\
$b_1$ &  0.012$_{0.04}^{9 \times 10^{38}}$  &  
5$\times 10^{65}$ $_{5 \times 10^{65}}^{1 \times 10^{117}}$ &   6$\times 10^{57}$ $_{6 \times 10^{57}}^{4 \times 10^{115}}$ & 1.2$_{-0.3}^{+0.5}$ \\
$b_2$ &  9$\times 10^{-141}$ $_{9 \times 10^{-141}}^{1 \times 10^{-38}}$  &   2$\times 10^{-102}$ $_{2 \times 10^{-102}}^{7 \times 10^{-11}}$ &   9$\times 10^{-94}$ $_{9 \times 10^{-94}}^{0.01}$ & 0.30$_{-0.04}^{+0.04}$  \\
\hline                                   
\end{tabular}
\end{table*}
%

\subsection{TDE fitting \label{tdefitting}}

To test whether the variability observed in AT\,2021hdr might be related to a TDE process, we fitted the ZTF data with the following decay model:

\begin{equation}
    \centering    
    Flux = C*t^{-D} ,
\end{equation}

\noindent where $Flux$ is the value of the difference flux (in $\mu$Jy), $C$ the amplitude, $t$ the normalized time and $D$ the power law decay exponent.

 \begin{figure}
   \centering
   \includegraphics[width=10cm]{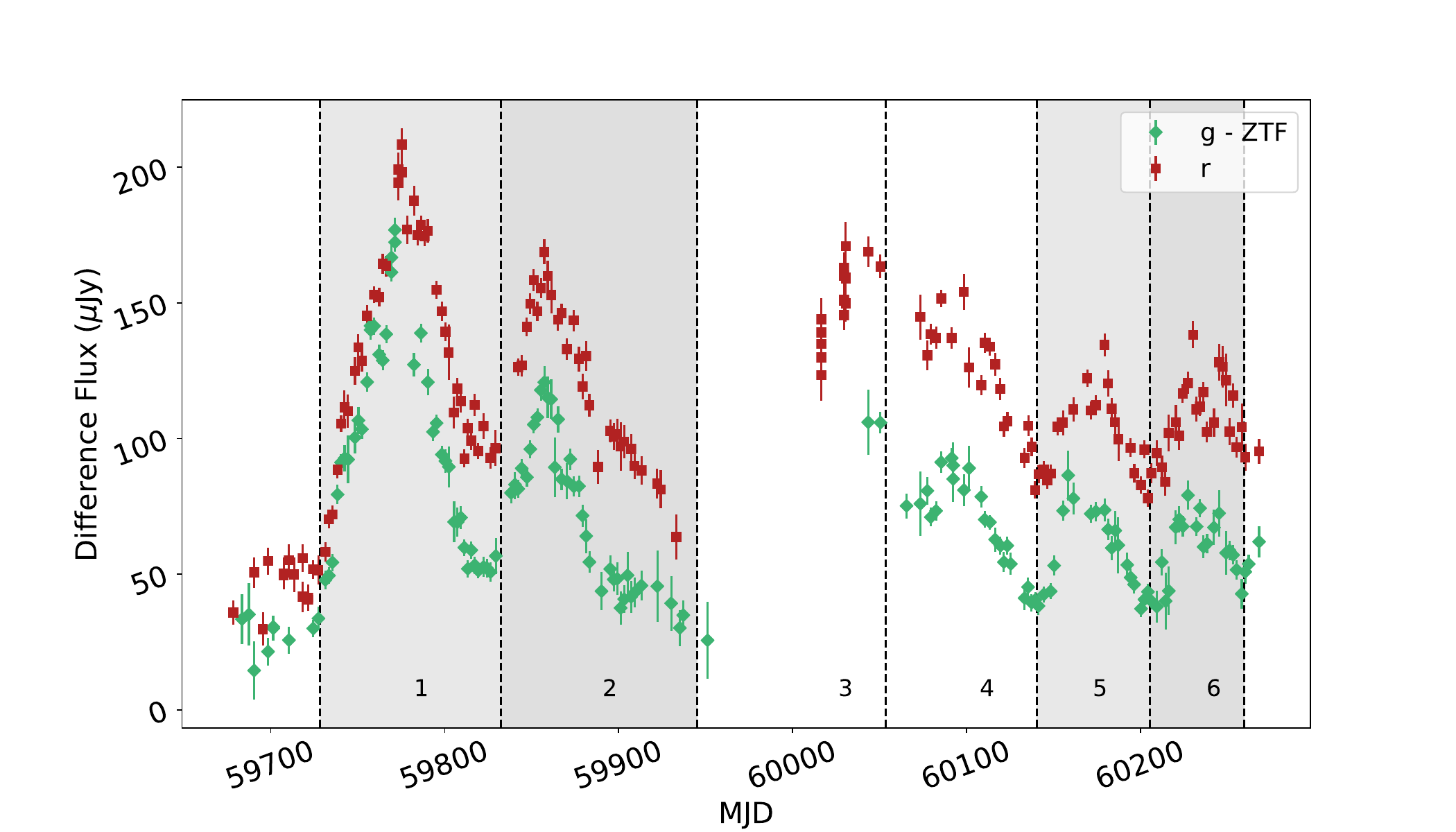}
      \caption{Zoom in to the ZTF light curve of AT\,2021hdr in the g and r bands. The dashed lines represent the dates that were used to define the individual peaks. The gray background represent the peaks that were used for the analysis of the individual peaks (see text).
              }
         \label{fig_tdefittingtotal}
   \end{figure}

A zoom in to the ZTF light curve of AT\,2021hdr is presented in Fig. \ref{fig_tdefittingtotal}. 
First, we divided the light curve of AT\,2021hdr into individual peaks with the aim of fitting each of them independently. 
The dashed lines in Fig. \ref{fig_tdefittingtotal} represent the dates that were used to define the individual peaks, that are named by numbers.
We used only the peaks that have a beginning and an end, i.e., peaks 1, 2, 5, and 6 (marked with a gray background in Fig. \ref{fig_tdefittingtotal}). We prefer to not use peaks 3 and 4 because it is unsure whether these are one or two peaks, and the source was behind the Sun during the rise of peak 3.

\begin{figure*}
   \includegraphics[width=8cm]{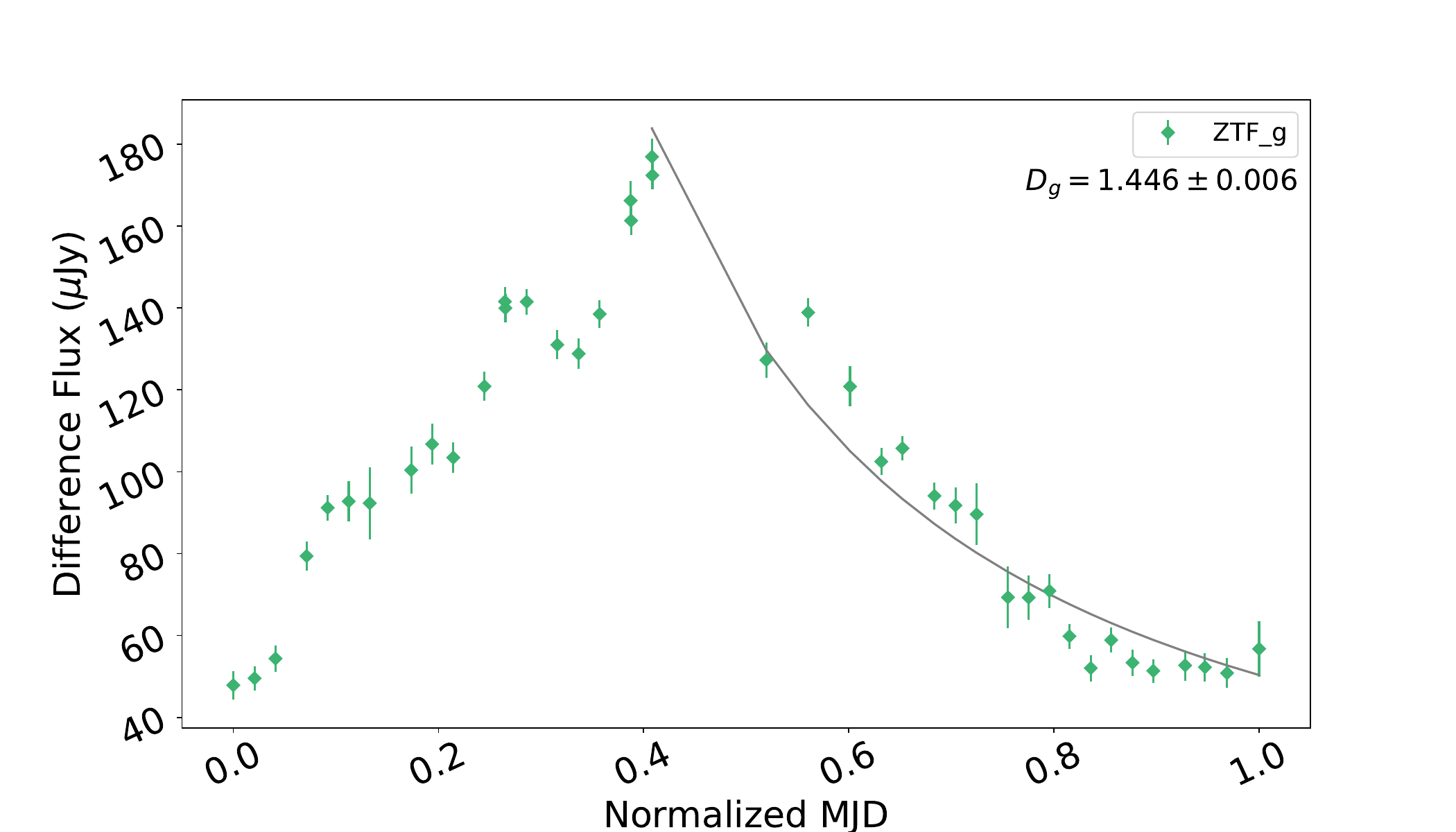}
   \includegraphics[width=8cm]{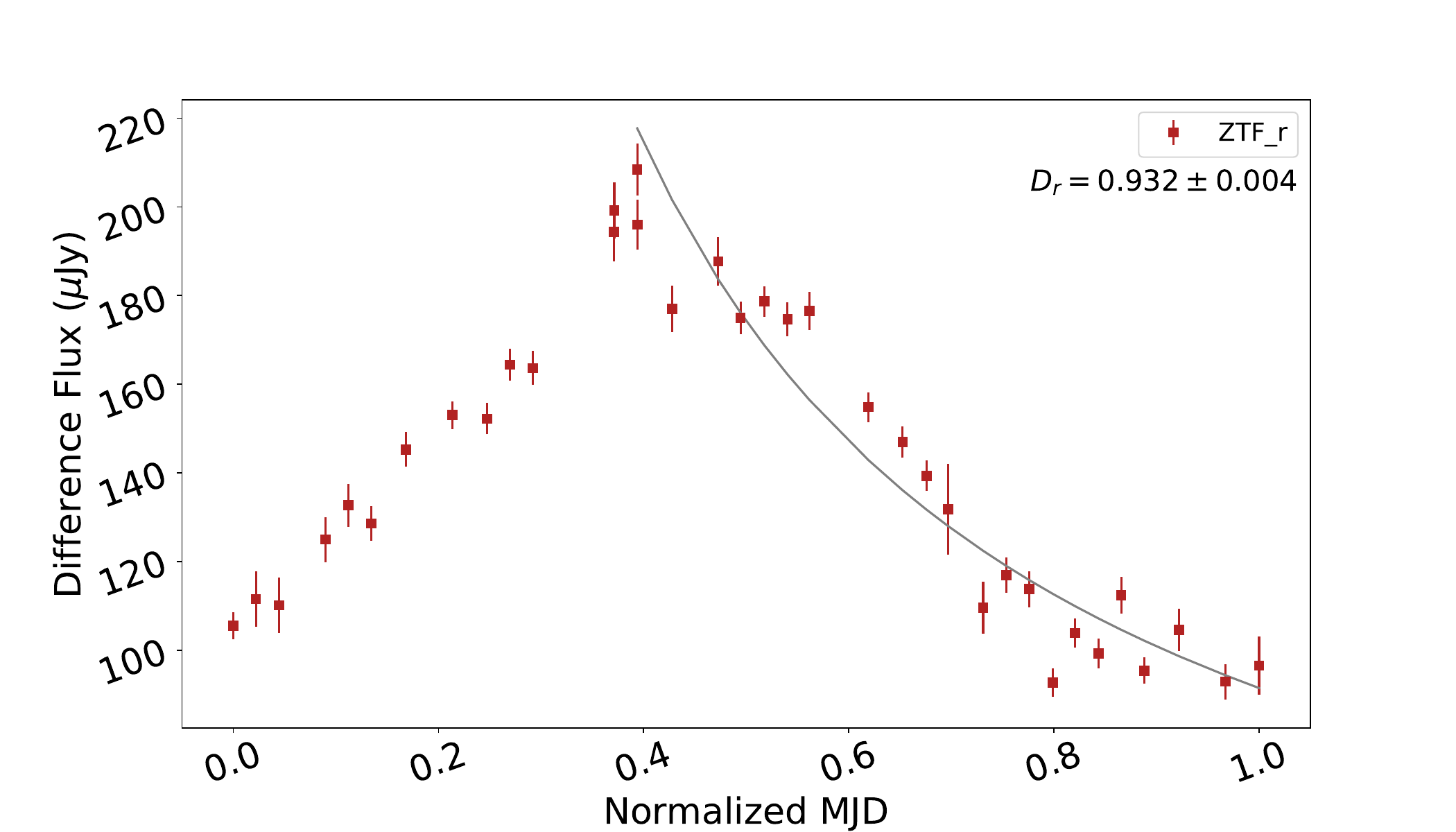}

    \includegraphics[width=8cm]{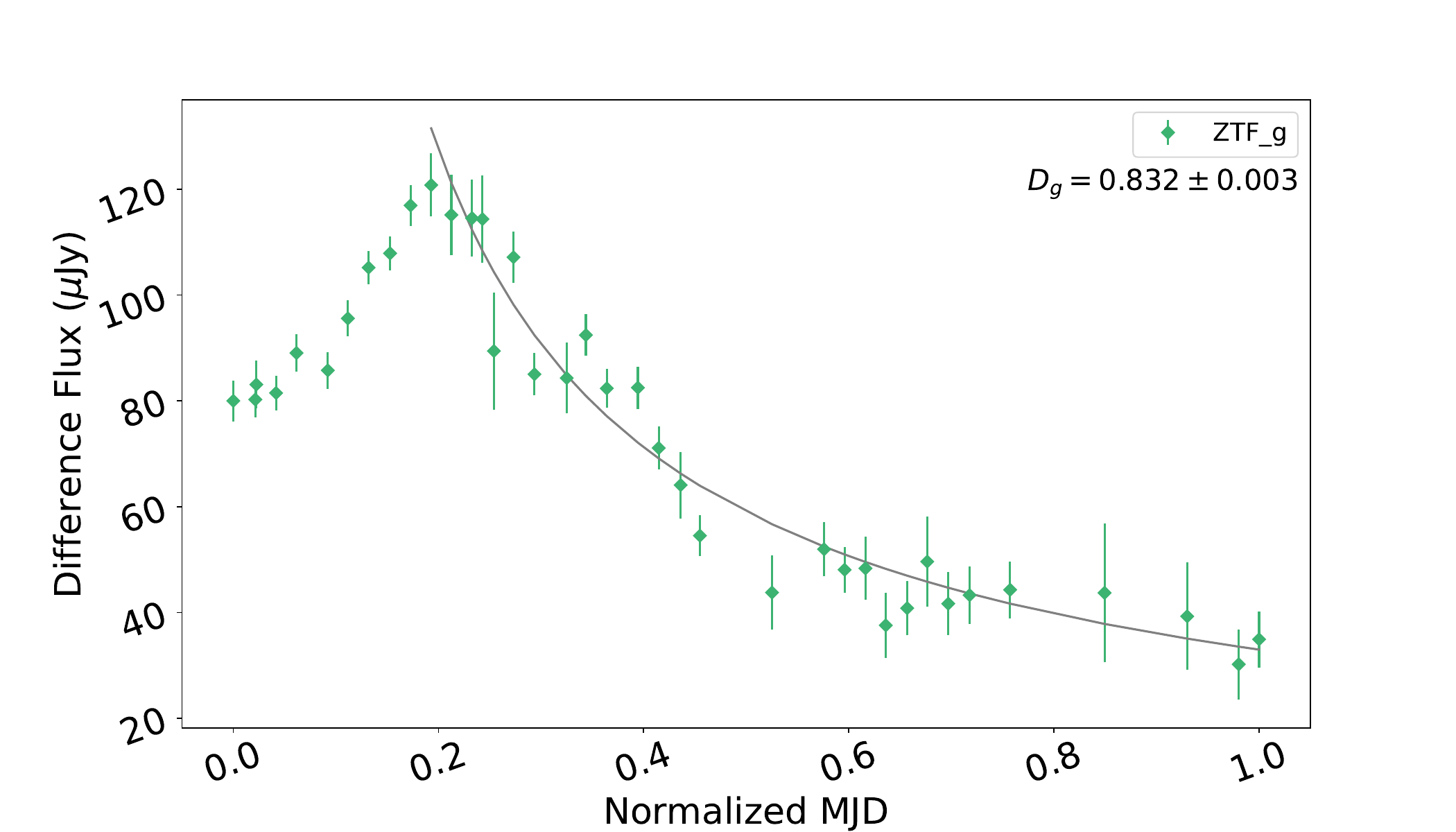}
   \includegraphics[width=8cm]{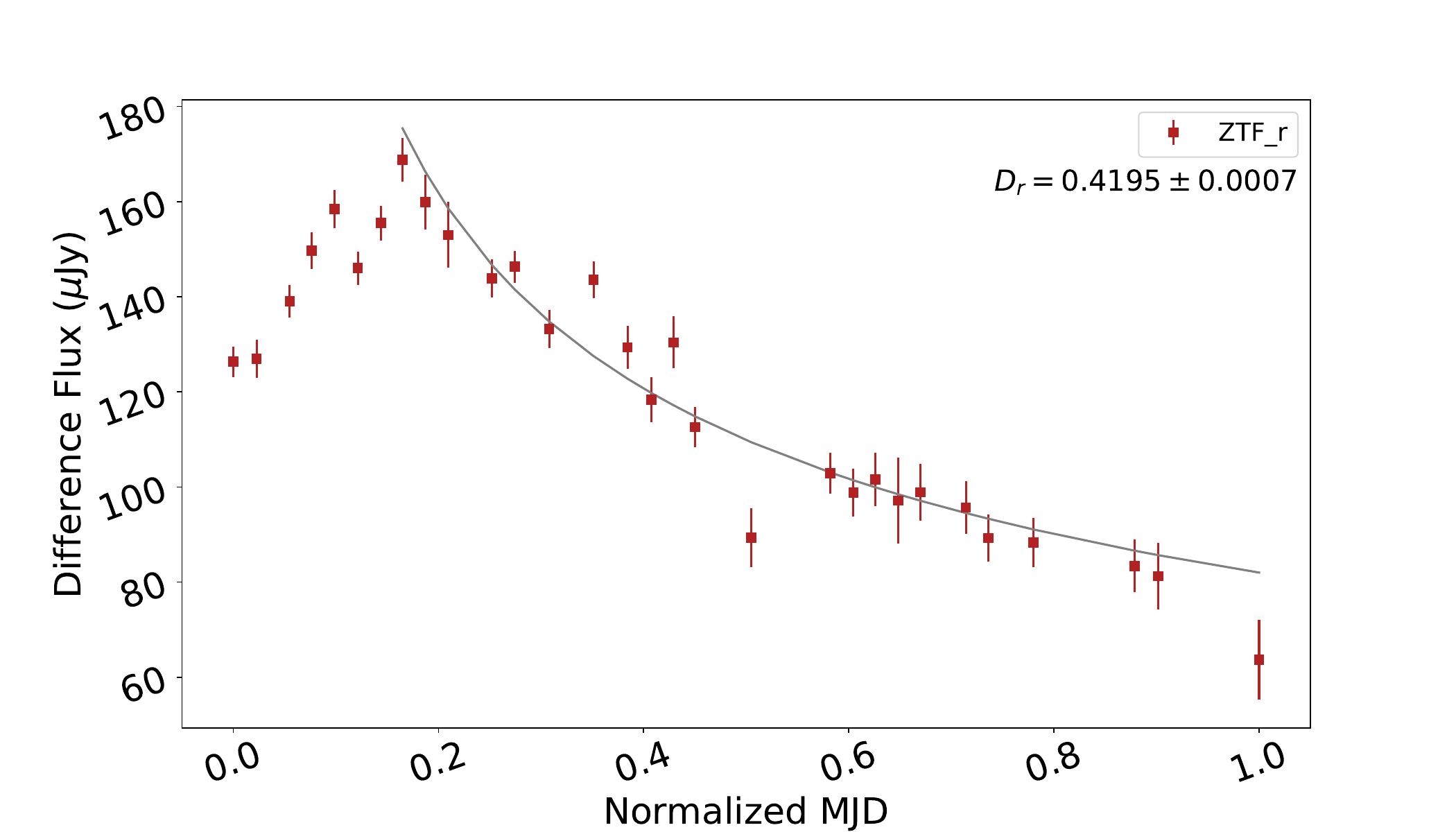}

  \includegraphics[width=8cm]{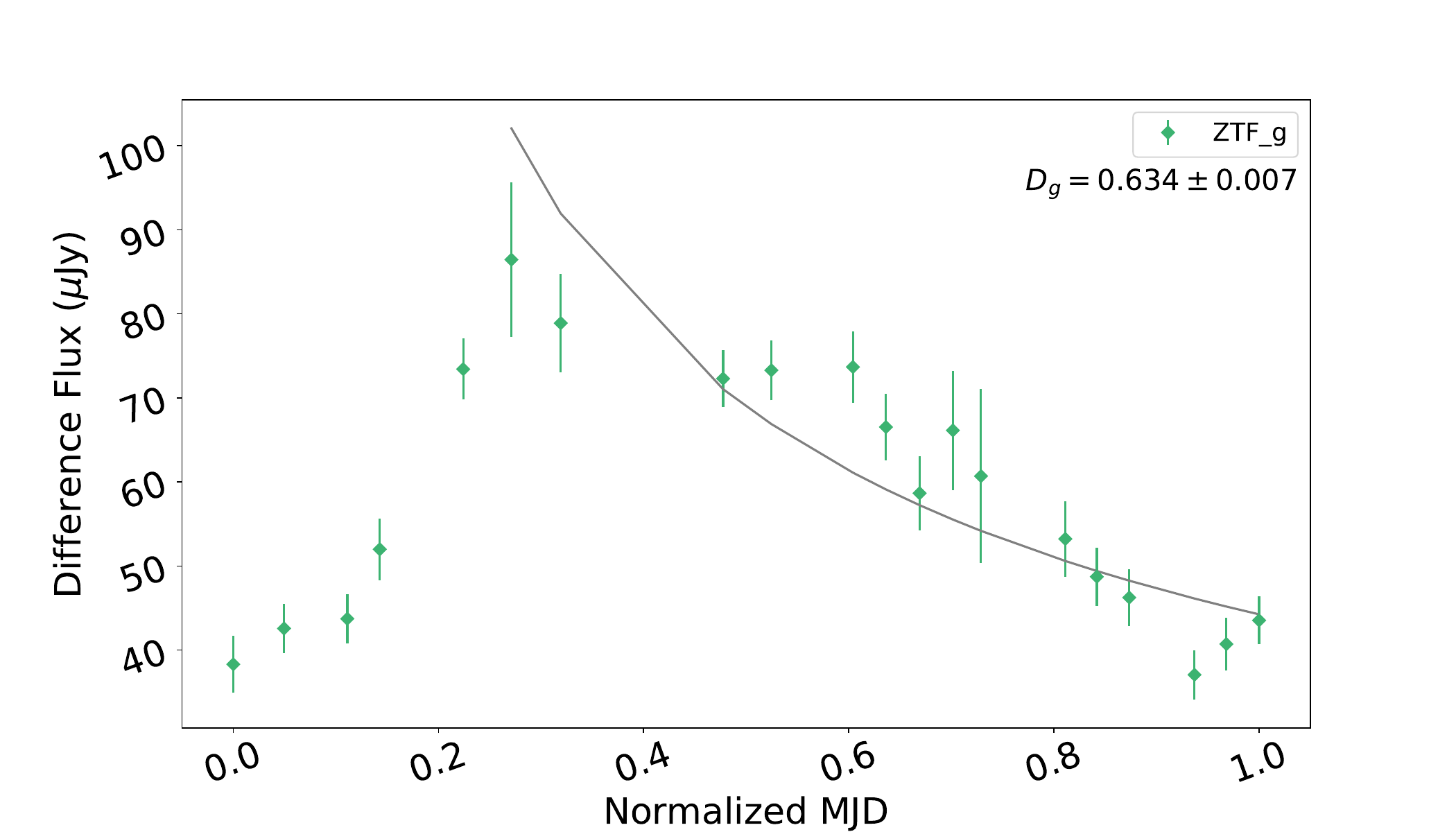}
   \includegraphics[width=8cm]{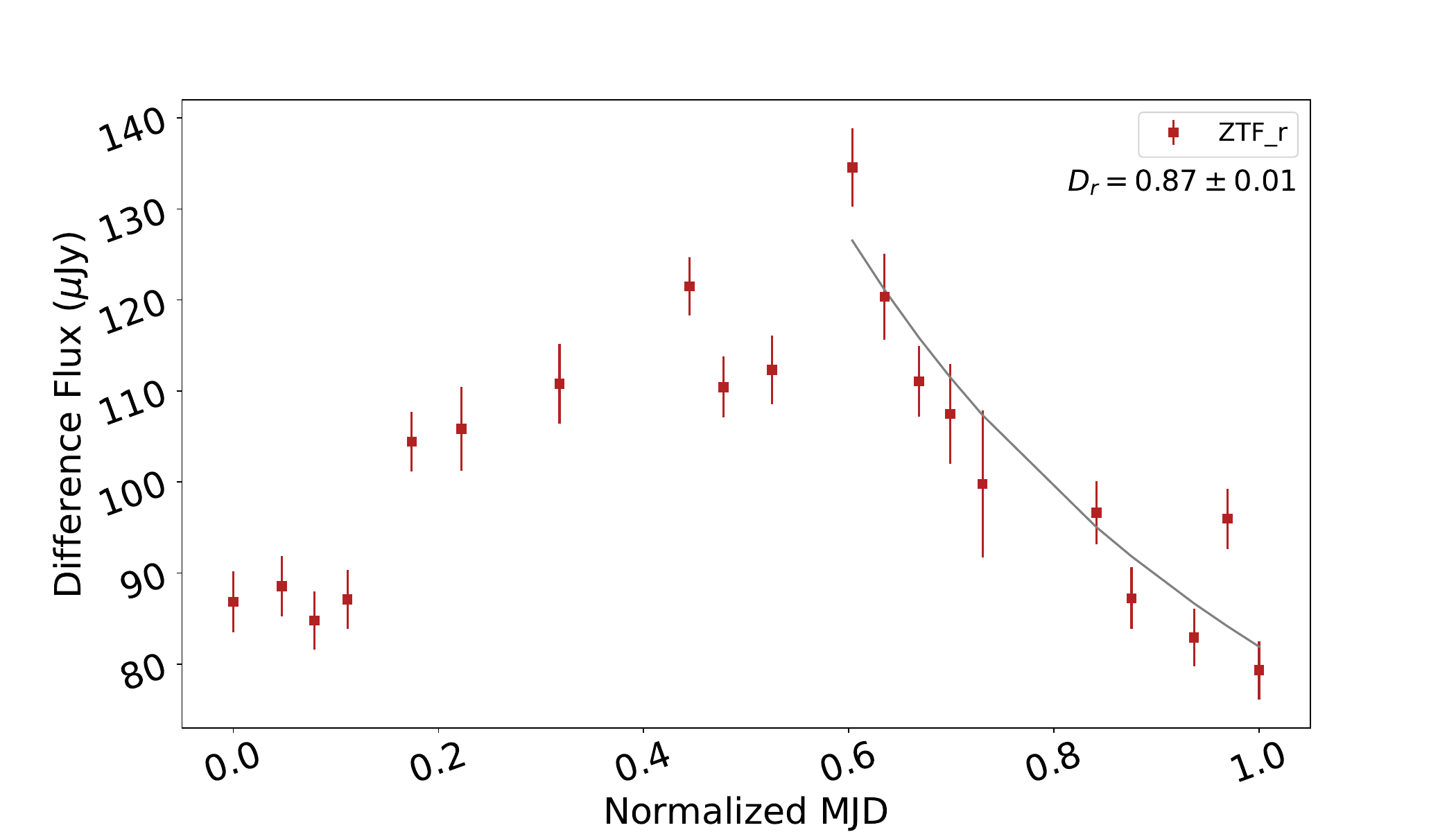}

   \includegraphics[width=8cm]{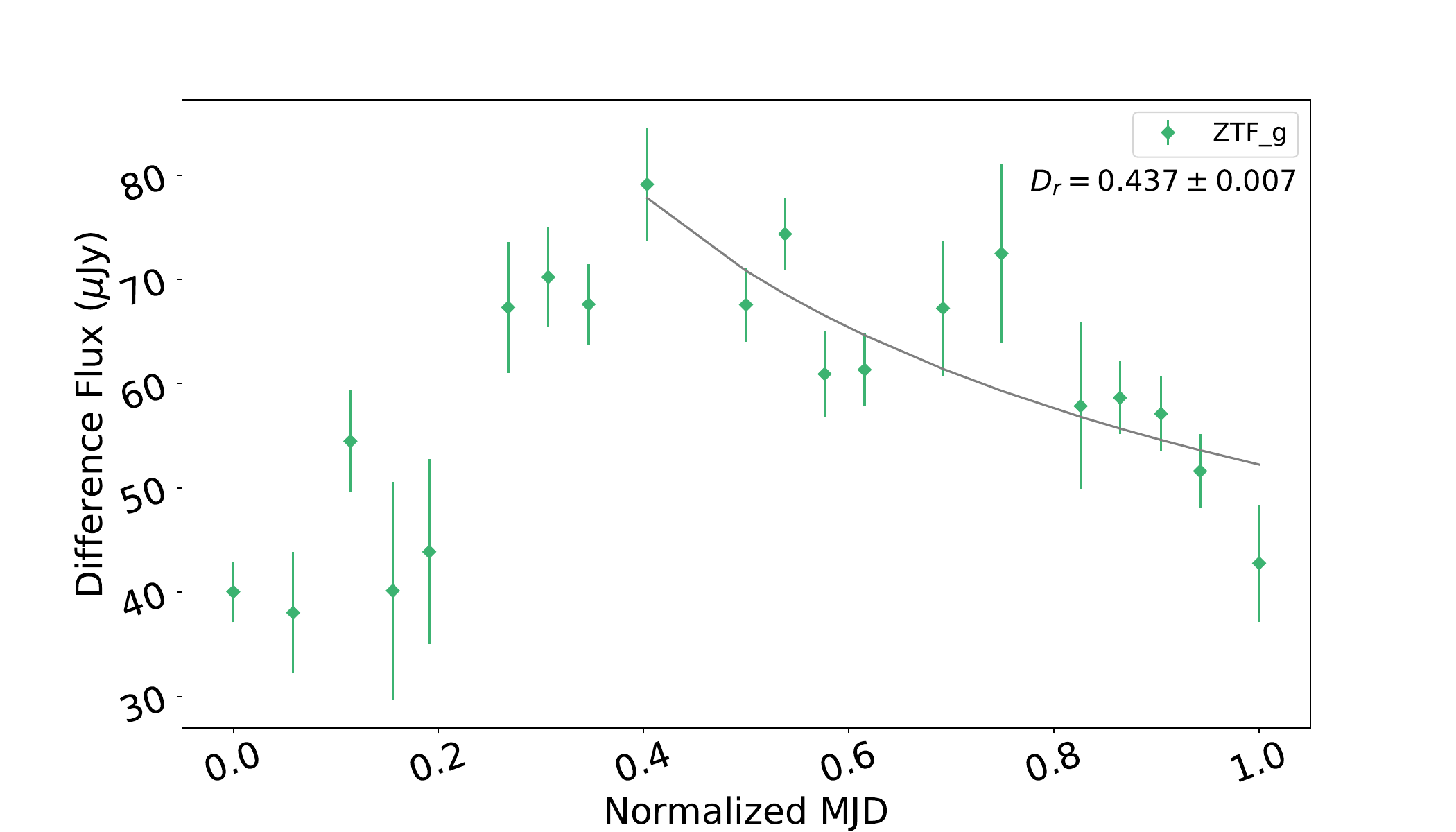}
    \includegraphics[width=8cm]{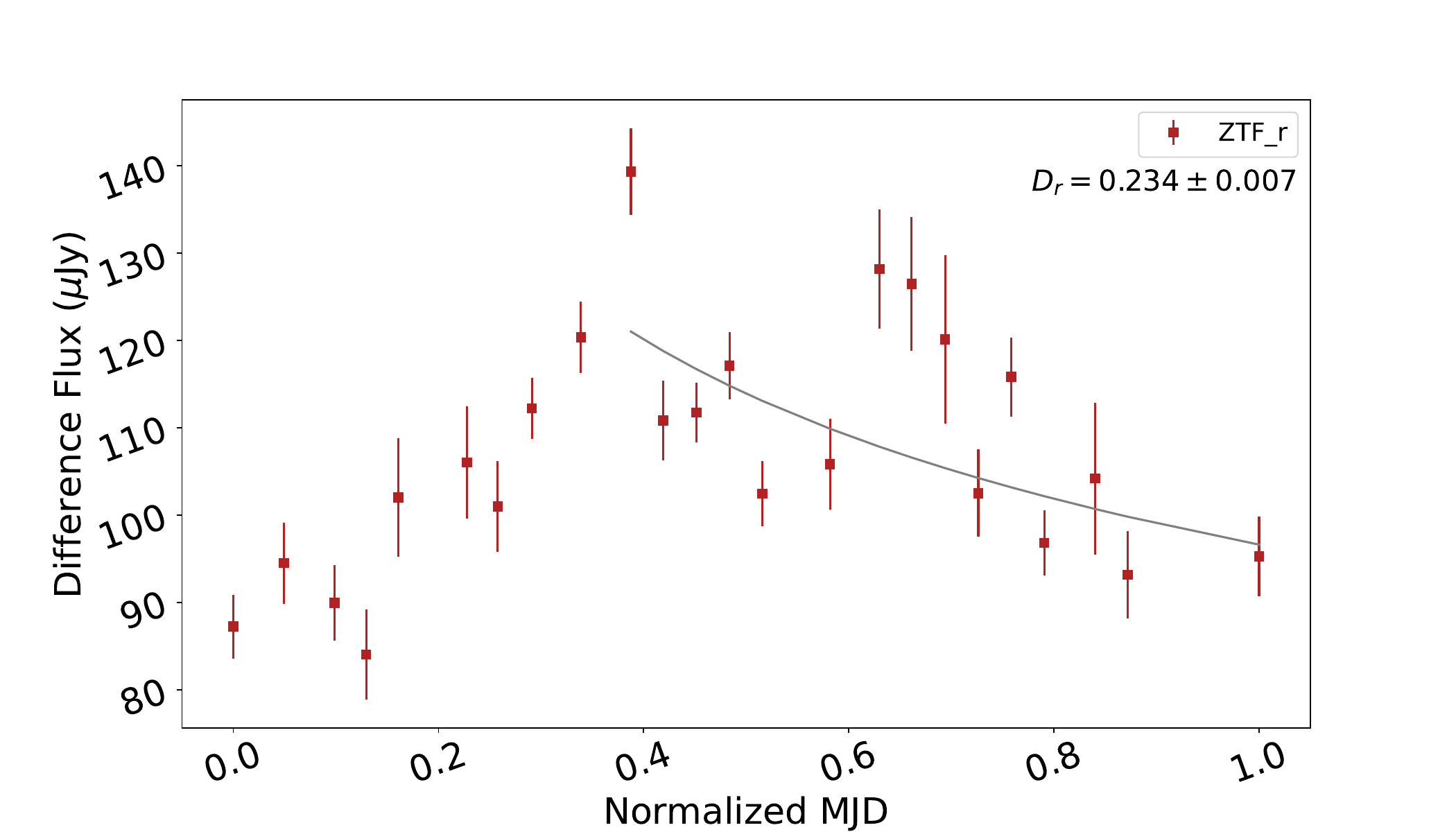}
      \caption{TDE fitting to each peak. The black lines represent a decay model with a free spectral index. From top to bottom, peaks 1, 2, 5, and 6 from Fig. \ref{fig_tdefittingtotal}. The legends include the ZTF band and the power law decay exponents.} 
         \label{fig_tdefittingind}
   \end{figure*}

The results of the fitting can be seen in Fig. \ref{fig_tdefittingind} and the decays for the g and r bands are in Table \ref{table:tdefit}. The mean values of the power law decay exponents are 0.8$\pm$0.4 for the g band, and  0.6$\pm$0.3 for the r band. 
The first decay exponent in the g band of the light curve of AT\,2021hdr is 1.445$\pm$0.006, but all the others are $<$1, much flatter than those expected for partial TDEs. This, together with the multiwavelength properties of AT\,2021hdr, makes it hardly to explain the nature of this transient as a partial TDE.

\begin{figure*}
   \centering
   \includegraphics[width=9cm]{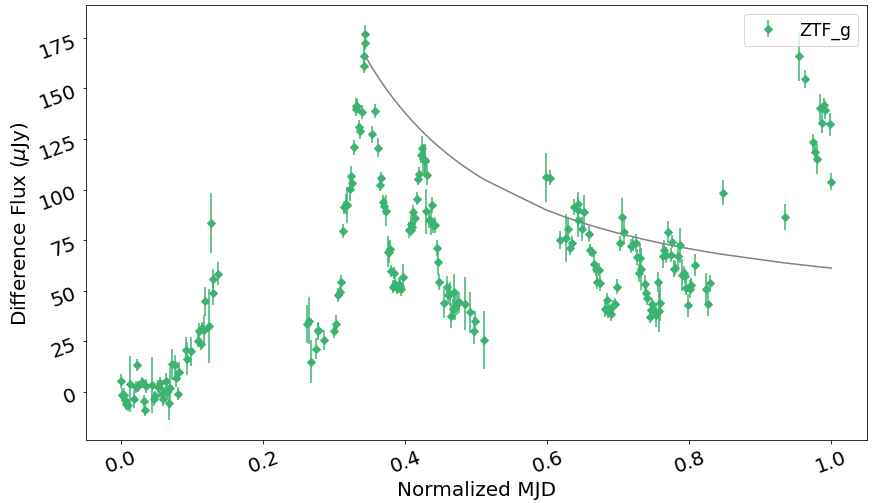}
   \includegraphics[width=9cm]{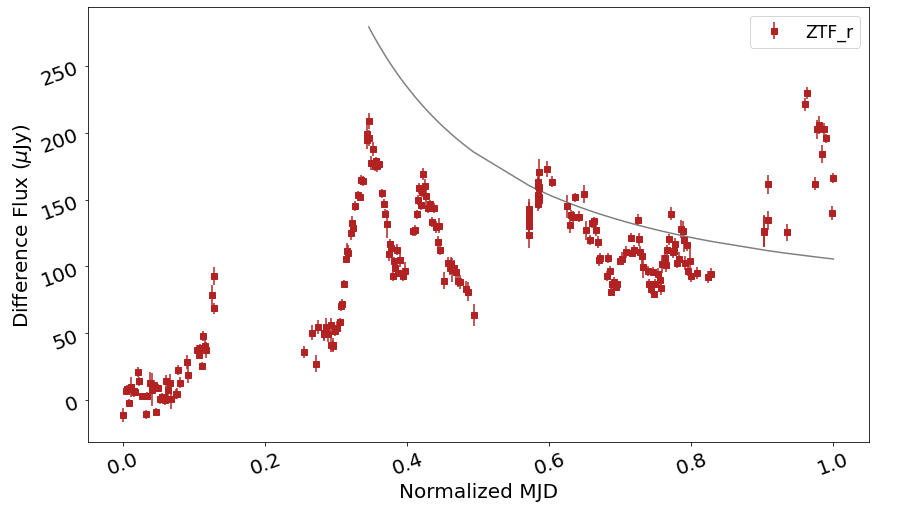}
      \caption{TDE fitting to the whole light curve in the g and r bands. The black line represents a decay model with spectral index 5/3 as expected for a TDE. The light curves include dates after MJD 59683 to consider only the oscillations. 
              }
         \label{fig_tdefitting}
   \end{figure*}

Then, we used all the peaks together to fit the same model but with a fixed D=5/3. We used data after MJD 59683 to consider only the oscillations.
With this model we want to test the scenario in which there is TDE by a BSMBH \citep[e.g.,][]{vigneron2018}. In this case we expect a sudden increase in accretion followed by an overall decay
with a power-law exponent of -5/3, but undergoing interruptions
due to the binary orbit. 
The result of this procedure is shown in Fig. \ref{fig_tdefitting} for g (left) and r (right) bands. From this plot it is obvious that the model is not a good fit for the data, mostly because of the newer bright peak  after 2024, discarding a TDE by a star in a BSMBH.

%
\begin{table}
\caption{TDE fitting. }             
\label{table:tdefit}      
\centering                          
\begin{tabular}{c c c}        
\hline\hline                 

Peak  &   $D_g$        &      $D_r$  \\
\hline
1 & 1.446$\pm$0.006 & 0.932$\pm$0.004 \\
2 & 0.832$\pm$0.003 & 0.4195$\pm$0.0007 \\
5 & 0.634$\pm$0.007 & 0.87$\pm$0.01 \\
6 & 0.437$\pm$0.007 & 0.234$\pm$0.007 \\
\hline                                   
\end{tabular}
\tablefoot{$D_g$ and $D_r$ correspond to the power law decay exponentials in the g and r bands, respectively, for each of the peaks defined in Fig. \ref{fig_tdefittingtotal}.}
\end{table}
%

\end{document}